\documentclass{jfm}

\usepackage{graphicx}
\usepackage{epstopdf, epsfig}
\let\proof\relax

\usepackage{amsmath,amsfonts,amssymb,amsthm,color,bbold,enumerate,stackengine}
\usepackage[round]{natbib}
\usepackage[utf8]{inputenc}
\usepackage[normalem]{ulem}
\usepackage{soul}
\usepackage[font=normal]{subcaption}
\usepackage{graphicx}

\usepackage{latexsym}
 \usepackage{cancel}
 \usepackage{multirow}
\usepackage[version=4]{mhchem}

\usepackage{hyperref}
\usepackage{mathtools}
\usepackage{titlesec}
\newcommand{\vect}[1]{\boldsymbol{#1}}
\newcommand\norm[1]{\left\lVert#1\right\rVert}

\shorttitle{Frequency selection in a gravitationally stretched capillary jet}
\shortauthor{I. Shukla, and F. Gallaire}

\title{Frequency selection in a gravitationally stretched capillary jet in the jetting regime}
\author{Isha Shukla\aff{1} \and Fran\c cois Gallaire\aff{1} \corresp{\email{francois.gallaire@epfl.ch}}}

\affiliation
{\aff{1}Laboratory of Fluid Mechanics and Instabilities, \'{E}cole Polytechnique F\'{e}d\'{e}rale de Lausanne, Lausanne, CH-1015, Switzerland}
\begin{document}
\maketitle
\begin{abstract}
A capillary jet falling under the effect of gravity continuously stretches while thinning downstream. We report here the effect of external periodic forcing on such a spatially varying jet in the jetting regime. Surprisingly, the optimal forcing frequency producing the most unstable jet is found to be highly dependent on the forcing amplitude. Taking benefit of the one-dimensional Eggers \& Dupont (J.~Fluid Mech.,~vol. 262,~1994,~205-221) equations, we investigate the case through nonlinear simulations and linear stability analysis. In the local framework the WKBJ formalism, established for weakly non-parallel flows, fails to capture the nonlinear simulation results quantitatively. However in the global framework, the resolvent analysis supplemented by a simple approximation of the required response norm inducing breakup, is shown to correctly predict the optimal forcing frequency at a given forcing amplitude and the resulting jet breakup length. The results of the resolvent analysis are found to be in good agreement with those of the nonlinear simulations.
\end{abstract}
Pele's hair, which are thin strands of volcanic glass formed in the air during the fountaining of the molten lava, is an impressive example of the stretching ability of highly viscous fluids. Named after Pele, the Hawaiian goddess of volcanoes, a single strand with a diameter of less than 0.5 mm, can extend up to a length of $2\,$m (\citealt{shimozuru1994physical,eggers2008}). If such viscous strands are pinned at one end, as in the case of honey dripping from a spoon under its own weight, gravity acts as the stretching tool for the viscous fluid producing very thin and stable liquid threads \citep{senchenko2005shape,Javadi2013}. The cross-section of such threads varies continually, as the jet accelerates downstream in the direction of gravity, before breaking into drops.  

Physically, the breakup of the jet into drops begins with the excitation of a temporally or spatially amplifying suitable mode due to weak external disturbances. In practice, this weak agitation is usually imposed by controlled harmonic perturbation, either from within or at the outlet of the nozzle, to generate spatially amplifying waves leading to jet breakup. In this direction, the primary objective of this paper is to evaluate the response of an incompressible jet falling in presence of gravity, to externally imposed harmonic perturbations characterised by a fixed frequency and amplitude, and to find the optimal forcing which generates the most amplified response. 

The external forcing is vital in the production of controlled micron-sized droplets, a feature essential to several application as in inkjet printers (\citealt{basaran2002small,wijshoff2010dynamics,basaran2013nonstandard}), pharmaceuticals \citep{bennett2002targeting} and powder technology \citep{van2013new}, to name a few. In view of the limitations linked to the fabrication of such small droplets, most of the devices used for the drops production depend on the generation of highly thin liquid threads whose diameters are several orders smaller than the nozzle diameter. Some common methods for producing such threads use tangential electrical stresses (in electrospinning devices, \citealt{doshi1995electrospinning, loscertales2002micro}), using outer co-flows \citep{marin2009generation} or a rotating spinerett (in fibre spinning applications \citep{pearson1969spinning}). \cite{Rubio} showed an alternative method for producing highly elongated jets through the use of gravity, in which the mass conservation of the liquid jet forces its thinning as the liquid accelerates downstream. 

The breakup of a liquid thread into drops, governed by the relative strength of the surface tension effect over the viscous and the inertial effects, was first explained by \cite{plateau1873} and \cite{rayleigh1879capillary} for a uniform column of fluid. 
What adds complexity to the well understood viscous jet breakup mechanism is the presence of gravity which significantly stretches the base flow shape. The stability of the such spatially varying gravity jets should ideally be examined using the global stability analysis and by including the non-parallel effects of the base flow. A similar difficulty linked to the non-parallel nature of the flow results from the adaptation of the flow from a wall bounded flow within the nozzle to a free jet \citep{sevilla2011effect}. Turning back to falling jets stretched by gravity, \citet{sauter2005stability} were the first to approach the problem theoretically by defining a linear global mode that correlated with the self sustained oscillations of the falling jet, observed during the jetting (globally stable) to dripping (globally unstable) transition. The work of \citet{sauter2005stability} was extended by \citet{Rubio} experimentally and theoretically by increasing the range of liquid viscosities and nozzle diameters. Additionally they retained the entire expression of the curvature term for the formulation of their stability analysis, a feature that helped them to accurately predict the critical flow rate for the stability transition and the oscillating mode compared to the previous author. However, none of these studies predicted the jet stable length as a function of the flow rate and fluid properties, a question which was pursued by \citet{Javadi2013} experimentally and theoretically. 

More recently, \citet{le2017capillary} determined theoretically the stable jet length, wavelength at breakup and resulting drop size due to the most dangerous perturbation applied either at nozzle exit or affecting the jet all along its length for different jet viscosities. Their analysis accounted for both the base state deformation and modification of local instability dispersion relation as the jet thins in the direction of gravity. Notably, extending the work of previous authors  (\citealp{tomotika1936breaking, frankel1985stability, leib1986generation,frankel1987influence, senchenko2005shape, sauter2005stability,Javadi2013}) they used the local plane wave decomposition (WKBJ approximation) for their analysis. However, the gain resulting from a perturbation was computed by considering only the exponential (e) terms of the WKBJ approximation. Additionally, an \textit{ad hoc} spatial gain of $\text{e}^7$, of the linear perturbations was assumed to be sufficient for breakup. Thus the level of noise was considered fixed for all the theoretical analysis. 

In this paper, we go beyond the global stability analysis of the gravity jets, and always operate in the stable regime where the jet behaves inherently as an amplifier. Precisely, we look at the receptivity of the jet in this regime to external perturbations, through nonlinear simulations and resolvent analysis with the aim of finding the optimal forcing which results in the most amplified disturbance. Unlike \citet{le2017capillary}, we consider an external forcing characterised by different amplitudes. The effect of forcing amplitude on the breakup length of very high speed jets has been numerically analysed by \citet{hilbing1996droplet}. However, a clear understanding of its effect in the case of spatially varying jets is still missing. Our analysis exemplifies the effect of forcing amplitude on the breakup length and the optimal forcing frequency. We also investigate the jet response using the WKBJ approximation and assess its validity for the spatially varying gravity jet. Our entire study is based on the slender-jet approximation \citep{eggers1994} of the Navier-Stokes equation for an axisymmetric jet. The reduced one dimensional (1D) model has turned out to be extremely valuable for realistic representation of jets (\citealt{ambravaneswaran2002drop,Hoeve2010}) by accurately capturing the jet interface close to the breakup as well as the formation of `satellite' drops. During the final stage of this work, we became aware of the work of \citep{LizziPhD} who has compared a resolvent analysis to experimental results and also revealed the dependance of the optimal frequency and breakup length as a function of time-harmonic forcing or noise amplitude.  

The paper is structured as follows. Section \ref{math_form} describes the governing equations. \S  \ref{chap3_dns} discusses the nonlinear simulations where the results are detailed in \S  \ref{DNS_results}. The local stability analysis of the gravity jet is performed in \S   \ref{LinStab} where we compare the jet response using  stability analysis in \S   \ref{spG1} and the WKBJ formulation in \S   \ref{sec_wkb}. We then operate in the global framework in \S  \ref{global} where the significance of the resolvent analysis is elucidated in \S  \ref{resol_ana}. We show that the resolvent analysis is self-sufficient in predicting the optimal forcing frequency and the breakup length as obtained through the nonlinear simulations. Finally, we apply a white noise disturbance on the jet inlet to explore its behaviour in comparison to the expected response to the optimal forcing in \S \ref{WhiteNoise}. The conclusion and some perspectives related to the present work are summarised in \S \ref{conc}.

\section{Mathematical formulation} \label{math_form}
We consider an axisymmetric viscous jet falling vertically from a nozzle under the effect of gravity $g$. At the nozzle outlet, the jet has a fixed radius $\bar{h}_0$ and velocity $\bar{u}_0$. The surrounding medium is considered evanescent and is neglected. The density, dynamic viscosity and surface tension of the jet are denoted by $\rho$, $\mu$ and $\gamma$, respectively. 

The behaviour of the jet is analysed using the leading-order one-dimensional mass and momentum equations, derived by \cite{eggers1994}. The dimensionless form of the equations, obtained by chosing $\bar{h}_0$ as the characteristic length scale, the inertial time $\tau_i=\sqrt{\rho \bar{h}_0^3/\gamma}$ as the characteristic time scale and  $\gamma/\bar{h}^2_0$ as the pressure scale, are written as,
\begin{subequations}\label{eq1_govEq}
\begin{align}
\frac{\partial h}{\partial t} &= -\frac{1}{2h}\frac{\partial}{\partial z} (h^2u), \\
\frac{\partial u}{\partial t} &= -u\frac{\partial u}{\partial z} - \frac{\partial p}{\partial z} + 3\mathit{Oh_{in}}\left(2\frac{\partial h}{\partial z}\frac{\partial u}{\partial z}\frac{1}{h} + \frac{\partial^2 u}{\partial z^2}\right) + \mathit{Bo_{in}},
\end{align}
\end{subequations}
\text{where, the dimensionless pressure $p(z,t)$ is expressed as,}
\begin{equation}\label{pr_dim}
\phantom{-}p = \left( \frac{1}{h\Big[1+\Big(\frac{\partial h}{\partial z}\Big)^2\Big]^{\frac{1}{2}}} - \frac{\frac{\partial^2 h}{\partial z^2}}{\Big[1+\Big(\frac{\partial h}{\partial z}\Big)^2\Big]^{\frac{3}{2}}} \right).
\end{equation}
In equation \eqref{eq1_govEq}, $h(z,t)$ and $u(z,t)$ represent the height of the jet interface and the velocity at the axial distance $z$. The system of equations \eqref{eq1_govEq} are governed by the dimensionless numbers Ohnesorge ($\mathit{Oh_{in}}$) and Bond ($\mathit{Bo_{in}}$) defined at the inlet. Ohnesorge, expressed as $\mathit{Oh_{in}}= {\mu}\big/{\sqrt {\rho \gamma \bar{h}_0}}$, relates the viscous forces to inertial and surface tension forces. The Bond (E\" otv\" os) number denoted by $\mathit{Bo_{in}} = \rho g{\bar{h}_0}^2\big/ \gamma$, measures the strength of the surface tension forces to body forces. A high $\mathit{Oh_{in}}$ or $\mathit{Bo_{in}}$ leads to a stabilised jet interface thus increasing the stability of the base flow. 

Using the associated characteristic velocity $\bar{h}_0/\tau_i$, the non-dimensional boundary conditions for the jet at nozzle inlet are reduced to:
\begin{subequations}\label{}
\begin{align}
h(0,t)&=1, \\
u(0,t)&=\sqrt{\mathit{We_{in}}}. 
\end{align}
\end{subequations}
Here, $\mathit{We_{in}}$ represents the Weber number defined at the nozzle inlet, $\mathit{We_{in}}= \rho \bar{h}_0 \bar{u}_0^2\big/\gamma$, and measures the ratio between the kinetic energy and the surface energy. 

The steady state form of the continuity equation (\ref{eq1_govEq}a) gives the relation between the steady state shape $h_{b}$ and velocity $u_b$ as, 
\begin{equation}\label{eq2_bs1}
h_b^2 u_b =Q =\sqrt{\mathit{We_{in}}}.
\end{equation}
Where $Q$ is the dimensionless flow rate, obtained from the nozzle conditions. This gives $u_b=\sqrt{\mathit{We_{in}}}/h_b^2$.
Using the relation \eqref{eq2_bs1}, the steady state momentum equation (\ref{eq1_govEq}b) reduces to, 
\begin{equation}\label{eq2_finalBS}
2Q^2h_b^{\prime} - h_b^{5}\mathit{C}^\prime + 6Q\mathit{Oh_{in}}(h_b{h_b^{\prime}}^2-h_b^2h_b^{\prime\prime})+h_b^5\mathit{Bo_{in}} =0,
\end{equation}
where derivatives are with respect to $z$ and $\mathit{C}$ is the jet interfacial curvature, expressed as,
\begin{equation}\label{eqcp}
 -h_b^2\mathit{C}^\prime= \frac{h_b^\prime}{[1+(h_b^\prime)^2]^{1/2}} +  \frac{h_b h_b^\prime h_b^{\prime\prime} + h_b^2h_b^{\prime\prime\prime}}{[1+(h_b^\prime)^2]^{3/2}}  -  \frac{3h_b^{2} h_b^\prime (h_b^{\prime\prime})^2}{[1+(h_b^\prime)^2]^{5/2}}.
\end{equation}
For the fixed nozzle inlet, equation \eqref{eq2_finalBS} is subject to boundary condition $h_b=1$ at $z=0$. Two more boundary conditions are needed to well define this differential problem of order three. However, exempting the jet tip from the base flow calculation gives us the liberty to impose a constant slope ($h_b^\prime=0$) and curvature ($h_b^{\prime\prime} =0$) at the exit of the jet. It should be noted that the boundary conditions applied at the jet exit should be treated as a way to close the differential problem rather than depicting physical boundary conditions. We made sure that these boundary conditions did not impact the overall base state solution by computing the solution over a large enough domain where the base state solution naturally converges to a solution with $h_b^\prime=0$ and $h_b^{\prime\prime}=0$.
\section{Nonlinear simulations} \label{chap3_dns}
The strength of a nonlinear simulation lies in its ability of capturing the exact response of the jet interface, including the shape close to the breakup point where the interface height $h$ approaches to a zero value. Often, the external forcing does not result in the breakup of fixed sized drops, rather the regular sized drops are followed by much smaller `satellite drops'.

Keeping this in view, we analyze the response of the jet in presence of an external forcing. We aim at finding the optimal forcing which results in the most unstable jet. The breakup length, which is the length of the stable jet between the nozzle and the breakup point, is chosen as the quantifier to compare the effect of different forcing, with the optimal forcing resulting in the shortest possible breakup length.

We begin with the description of the modified non-linear governing equations used for the simulations followed by the numerical scheme implemented to capture jet breakup. Finally we present the comparison of breakup characteristics of the jet for different inlet forcing . 
\subsection{Governing equations}
In order to remove the singularity in expression (\ref{pr_dim}) for the pressure, when $h(z,t)\to0$, we define the interface height $h(z,t)$ in terms of function $a(z,t)$ where $a=h^2$. The governing equations \eqref{eq1_govEq} thus transform into,
\begin{subequations}\label{eq3_govEq_dns}
\begin{align}
\frac{\partial a}{\partial t} &= -\frac{\partial}{\partial z} (au),\\
\frac{\partial u}{\partial t} &= -u\frac{\partial u}{\partial z} - \frac{\partial p}{\partial z} + 3\mathit{Oh_{in}}\Bigg(\frac{\partial}{\partial z}\Big(a \frac{\partial u}{\partial z}\Big)\frac{1}{a} \Bigg) + \mathit{Bo_{in}},\\
p &= \left( \frac{\Big(2-\frac{\partial^2 a}{\partial z^2}\Big)a + \Big(\frac{\partial a}{\partial z}\Big)^2}{2\Big(\frac{1}{4}{\bigg(\frac{\partial a}{\partial z}}\bigg)^2 + a\Big)^{\frac{3}{2}}}\right). 
\end{align}
\end{subequations}
The base state solution for the jet interface $(a_b=h_b^2)$ is obtained by solving equation \eqref{eq2_finalBS}. We model the external forcing on the jet by perturbing only the inlet velocity using a forcing of the form,
\begin{equation}\label{eq3_dnsForcing}
 u_f(0,t )={\rm I\!R}( \epsilon e^{\text{i}\omega t}),
 \end{equation} 
where $\epsilon$ represents the amplitude of the forcing and $\omega$ represents the angular forcing frequency. In presence of the forcing, the boundary conditions at the inlet are modified to $a(0,t)=1$ and $u(0,t)=\sqrt{\mathit{We_{in}}}+u_f$.  No boundary conditions are defined at the other extremity of the domain close to the tip. Nonetheless, a special treatment is applied for the tip as explained in the next section. 

\subsection{Numerical scheme} \label{Chap3_NumScheme}
\label{NumericalScheme}
The governing equations \eqref{eq3_govEq_dns} are first discretised in space, after which the resulting ordinary differential equations (ODE) are integrated in time. Diffusion terms are evaluated using second-order finite differences, with a central scheme for intermediate nodes and a forward or backward scheme for boundary nodes. Advection terms are obtained using a weighted upwind scheme inspired by \cite{spalding1972novel} hybrid difference scheme. Unlike the latter, which approximates the convective derivative using a combination of central and upwind schemes, we evaluate the derivative based on a combination of forward and backward finite differences. An advection term $dA/dz$ is evaluated at node $i$ as
\begin{equation}\label{eqNb1}
\left( \frac{dA}{d z} \right)_i = \beta \left( \frac{dA}{d z} \right)_{i,b} + (1-\beta) \left( \frac{dA}{d z} \right)_{i,f},
\end{equation}
where indices $b$ and $f$ refer to the backward and forward finite difference schemes, and $\beta$ is a weight coefficient that depends on the local value of velocity $u$ at node $i$ together with a parameter $\alpha$,
\begin{equation}\label{eqNb2}
\beta = \frac{\tanh(\alpha u_i)+1}{2}.
\end{equation}
For the range of feed velocities considered in this study, numerical stability was always ensured by using a 10-point stencil. Thus, the backward difference term relies on a stencil that spans nodes $i-5$ to $i+4$, and the forward difference term employs nodes $i-4$ to $i+5$. For large enough downstream or upstream velocities, $\beta$ will tend to $1$ or $0$ respectively; hence \eqref{eqNb1} reduces to a regular upwind difference scheme. For smaller velocity magnitudes in between, \eqref{eqNb1} produces a weighted combination of backward and forward differences. In our simulations, we choose $\alpha = 50$ so that the transition between the backward and forward difference schemes mostly occurs when $|u| < 0.05$. Finally, advection terms at nodes close to the boundary are evaluated based on the values of the closest 9 adjoining nodes. 

After obtaining all spatial derivatives, the resulting ODEs are integrated using the MATLAB solver \texttt{ode23tb}, which implements a trapezoidal rule and backward differentiation formula known as TR-BDF2 \citep{bank1985transient}, and uses a variable time step to reduce the overall simulation time. 

The numerical domain $L$ is taken sufficiently large to capture the breakup of the jet. The jet interface is initialised by the solution of \eqref {eq2_finalBS} obtained numerically with the MATLAB \texttt{bvp4c} solver. The validation of the numerically obtained base state solution is presented in Appendix \ref{app_1g}. It should be noted that the steady state is implemented only for a part of the numerical domain and the interface is initialised to 0 for the remaining part. The axial span of the base state solution does not affect the quasi-steady jet characteristics, which are the focal point of our numerical analysis. A validation for the same is presented in Appendix \ref{app_3g}.

At every time step, the solution is evaluated for three conditions: (\textit{i}) \textit{Pinch-off (breakup):} It is defined as when the value of $a$ passes below a threshold value of $10^{-5}$. The corresponding time $\mathit{t_{po}}$ is saved and the position of the jet tip is updated as $\mathit{N_{tip} = N_{po}}$, where $\mathit{N_{po}}$ is the pinch-off location. The solution for $a$ and $u$ beyond $\mathit{N_{tip}}$ is set to zero. For subsequent time steps, $\mathit{N_{tip}}$ has two possibilities -- it can either advance or recede, which requires the following two conditions. (\textit{ii}) \textit{Advancing jet:} The values of $a$ at nodes $\mathit{N_{tip}-1}$ and $\mathit{N_{tip}}$ are extrapolated to find $a$ at $\mathit{N_{tip}+1}$. If the extrapolated value is larger than a predefined value of $5 \cdot 10^{-3}$, the parameter $\mathit{N_{tip}}$ is incremented by 1, and $a$ and $u$ at the new $\mathit{N_{tip}}$ are assigned values extrapolated from its previous two neighbours. (\textit{iii}) \textit{Receding jet:} If the value of $f$ at $\mathit{N_{tip}}$ falls below a predefined value of $10^{-3}$, $a$ and $u$ at $\mathit{N_{tip}}$ are set to zero and the parameter $\mathit{N_{tip}}$ is reduced by 1. These three conditions enable the numerical integration of the governing equations in a way that captures accurately the breakup of the jet and the motion of the tip. A validation of the numerical scheme is presented in Appendix \ref{app_2g}. 
\subsection{Nonlinear simulation results}\label{DNS_results}

Using the numerical scheme presented in the previous section, nonlinear simulations were performed for a jet governed by equation \eqref{eq3_govEq_dns} for fixed inlet characteristics: $\mathit{Oh_{in}}=0.3$, $\mathit{We_{in}}=1.75$ and $\mathit{Bo_{in}}=0.1$. The jet inlet velocity is subjected to time harmonic forcing of the form given by equation \eqref{eq3_dnsForcing} with a fixed amplitude $\epsilon$ and for forcing frequency $\omega=[0.4-3.2]$

The simulations were run for a sufficiently long time to enter a permanent regime wherein the jet breaks up at regular intervals of time and at fixed axial location. In the quasi-steady regime, the breakup period $\mathit{\Delta T_{po}}$ is defined as the time difference between two consecutive breakups or pinch-offs and the breakup length $l_c$ as the stable length of the jet between the nozzle and the pinch-off location. We use $l_c$ as the quantifier to determine the stability of the jet to external forcing such that the most amplified disturbance caused by the optimal frequency $\mathit{\omega_{opt}}$ will compel the jet to have the shortest possible breakup length. 
\begin{figure}
\centering
\includegraphics[trim= 5 0 0 0,clip,width=1.01\linewidth]{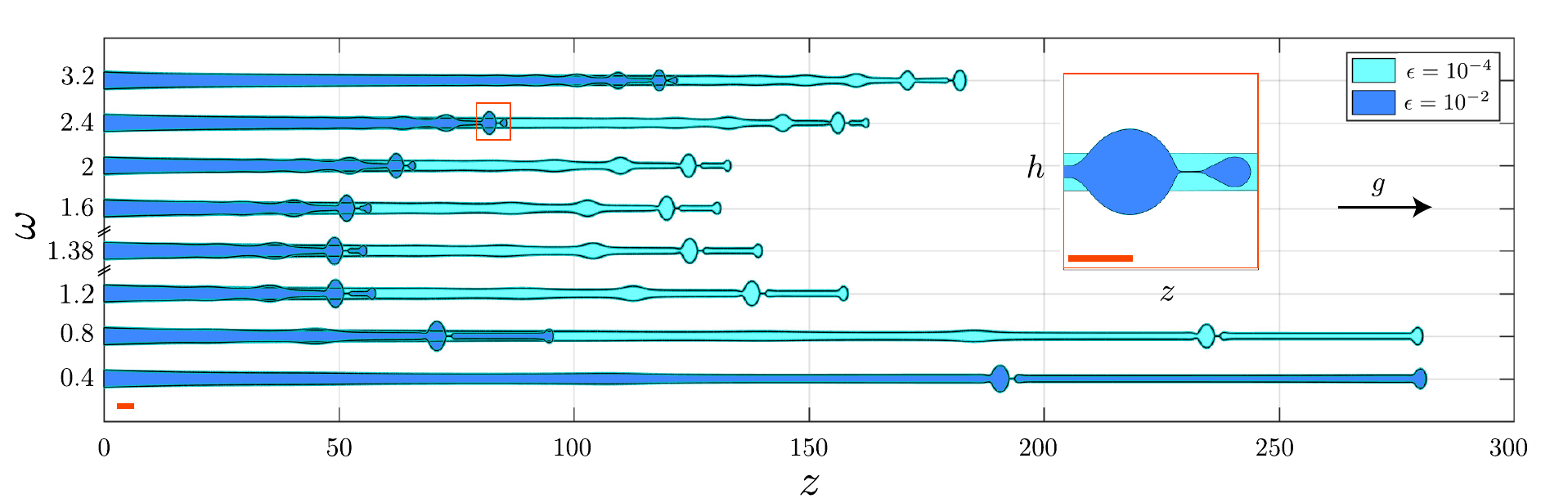}
\caption{The plot shows the jet intact shape along the axial direction $z$ for a gravity jet defined by $\mathit{Oh_{in}}=0.3$, $\mathit{Bo_{in}}=0.1$ and $\mathit{We_{in}}=1.75$ and perturbed by different inlet forcing frequencies $\omega$ and forcing amplitudes $\epsilon=10^{-2}$ and $\epsilon=10^{-4}$. For clarity, only the shape corresponding to the shortest breakup length for every frequency is plotted. The jets of two different colors represent the shape at \textit{approximately} the same forcing frequency but different forcing amplitudes $\epsilon$. We see that for $\epsilon=10^{-2}$, the the breakup length is the minimum for $\omega=1.38$ and for $\epsilon=10^{-4}$ for $\omega=1.68$.  The box in red shows the zoomed image of the jet close to a breakup highlighting the existence of a satellite and a main drop. The zoomed image has radial $h$ and axial $z$ dimensions drawn to the same scale each representing a dimensionless size of 6. The red bar in both the plots represents a dimensionless radial length scale of 2, which is also the size of the dimensionless nozzle diameter.} 
\label{fig:WN1b}
\end{figure}%

We begin our analysis for fixed amplitudes $\epsilon=10^{-2}$ and $\epsilon=10^{-4}$. The response of the jet due to different forcing frequency $\omega$ in the permanent regime for the two above mentioned amplitudes can be seen in Fig. \ref{fig:WN1b}. Jets enforced by the same disturbance amplitude at the inlet are represented by the same colour. For visual clarity we plot the response only for certain frequencies and for the jet shape pertaining to the shortest breakup length. First, Fig. \ref{fig:WN1b} clearly shows the existence of a main drop and a satellite drop for all the frequencies. Second, for $\epsilon=10^{-2}$ we conclude that the optimal forcing frequency is $\mathit{\mathit{\omega_{opt}}}=1.38$ because it manifests the jet to have the shortest $l_c$. Third, and most strikingly, we notice that for a lower forcing amplitude of $\epsilon=10^{-4}$, the optimal forcing increases to $\mathit{\omega_{opt}}=1.68$. Finally, at all forcing frequencies, the breakup length for jet with $\epsilon=10^{-4}$ is always larger than for $\epsilon=10^{-2}$.

\begin{figure}
\centering
\includegraphics[trim=120 0 70 0,clip,width=0.85\linewidth]{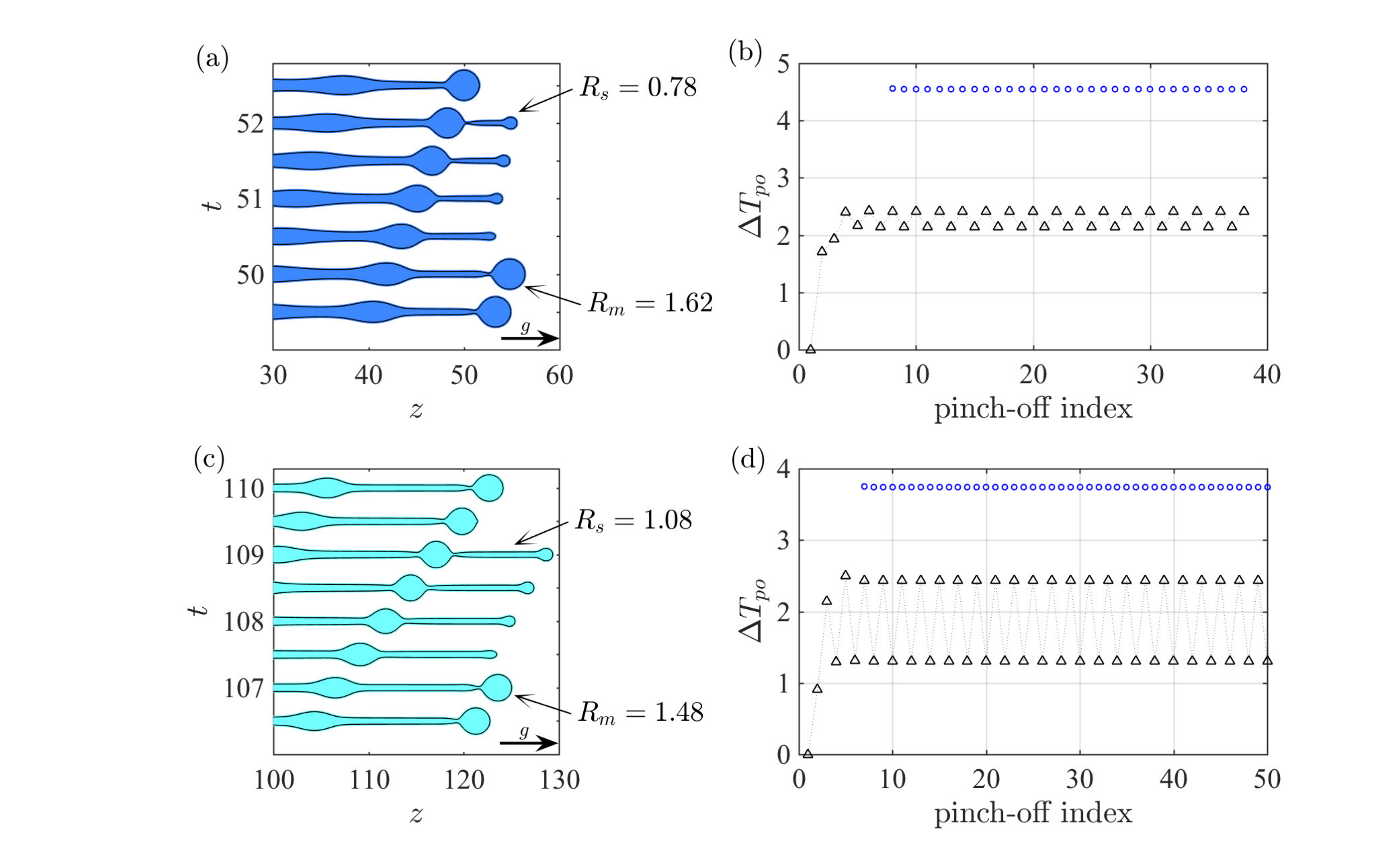}
\caption{Breakup characteristics for a gravity jet defined by $\mathit{Oh_{in}}=0.3$, $\mathit{Bo_{in}}=0.1$ and $\mathit{We_{in}}=1.75$ and perturbed with $\mathit{\omega_{opt}}$. Subplots (a) and (b) correspond to a forcing with $\epsilon=10^{-2}$ and $\mathit{\omega_{opt}}=1.38$. Subplots (c) and (d) refer to a forcing with $\epsilon=10^{-4}$ and $\mathit{\omega_{opt}}=1.68$. Subplots  (a) and (c) elaborate the interface profile at the time of breakup with the existence of a satellite drop after the main drop is released. The main and satellite drop radius for both the cases have been highlighted. The axial and radial dimensions of (a) and (c) represent the same length scale. The breakup period $\mathit{\Delta T_{po}}$ is represented in (b) and (d). The black triangle refers to the $\mathit{\Delta T_{po}}$ between the consecutive drops and the blue circle represents the one between two consecutive main (or satellite) drops. The breakup frequency $\omega_{po}$ is equal to 1.38 and 1.68 in (b) and (d) respectively.} 
\label{fig:WN1c}
\end{figure}

To investigate further the breakup characteristics for the amplitudes $\epsilon=10^{-2}$ and $\epsilon=10^{-4}$ due to $\mathit{\omega_{opt}}$, we plot the interface evolution in the permanent regime as shown in Fig. \ref{fig:WN1c}(a) and Fig. \ref{fig:WN1c}(c), where regular sized main drop formation is followed by the release of a satellite drop. For both the amplitudes, we see a distinct difference between the main and satellite drop radius. 

The breakup period $\mathit{\Delta T_{po}}$ resulting from the forcing imposed in Fig. \ref{fig:WN1c}a and \ref{fig:WN1c}c are plotted in Fig. \ref{fig:WN1c}b and \ref{fig:WN1c}d, respectively, where the black triangle represent the $\mathit{\Delta T_{po}}$ obtained for two consecutive pinch-offs whereas the blue circles denote $\mathit{\Delta T_{po}}$ obtained for two consecutive pinch-offs of the \textit{same group}, that is to say between two consecutive main (or satellite) drops. From the figure we conclude that even though the breakup period is the same for the group of main and satellite drops (as shown by blue circles) the time of formation of a satellite drop does not lie exactly midway between the time of formation of the main drops and vice versa. This results in obtaining two oscillating breakup periods (as shown by the black triangles). We further observe that the frequency of breakup ($\omega_{po}=2 \pi /\Delta T_{po}$) obtained using the breakup period for consecutive main (or satellite) drops responds to the externally applied forcing at the jet inlet with $\omega_{po}=1.38$ and $1.68$ for $\epsilon=10^{-2}$ and $10^{-4}$, respectively.

Finally, for the constant flow rate of the jet, the breakup period related to the consecutive pinch-off's is used for obtaining the drop radius for the satellite and main drops. We notice that at the optimal forcing frequency, the main drop radius $R_m$ decreases from 1.62 to 1.48 dimensionless units as $\epsilon$ reduces from $10^{-2}$ to $10^{-4}$. On the contrary, the satellite drop radius $R_s$ increases from 0.78 to 1.08 dimensionless units for $\epsilon=10^{-2}$ and $\epsilon=10^{-4}$ respectively. The longer intact jet length obtained for lower forcing amplitude $\epsilon=10^{-4}$ results in a larger downstream velocity close to the tip due to the presence of gravity. Eventually, it results in the formations of highly stretched satellite drops in comparison to the ones obtained for lower amplitude of $\epsilon=10^{-2}$ as seen in Fig. \ref{fig:WN1c}a and \ref{fig:WN1c}c. 
\begin{figure}
\centering
\includegraphics[trim=50 10 50 0,clip,width=1\linewidth]{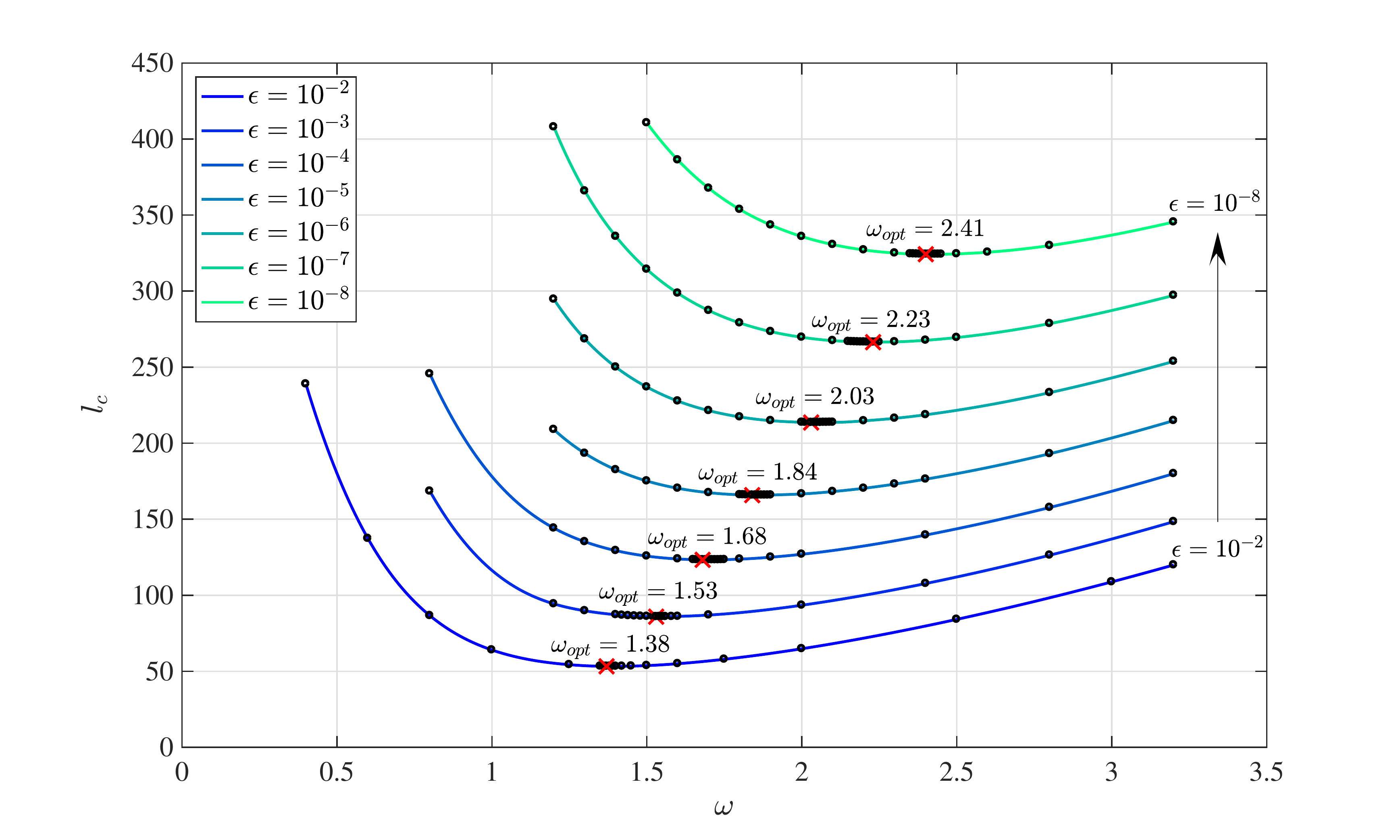}
\caption{The plot shows the breakup length $l_c$ as a function of forcing frequency $\omega$ for a gravity jet defined by $\mathit{Oh_{in}}=0.3$, $\mathit{Bo_{in}}=0.1$ and $\mathit{We_{in}}=1.75$. Each curve is indicative of a fixed forcing amplitude $\epsilon$. For a fixed $\epsilon$, the optimal forcing frequency related to the shortest $l_c$ is represented by a red cross. We observe that the optimal frequency increases as $\epsilon$ decreases and does not appear to saturate even for lower  amplitudes of $10^{-8}$. The black circles represent the data from numerical simulations.} 
\label{dns-epsilon1}
\end{figure}%

We now return to the most salient feature observed in Fig. \ref{fig:WN1b}, where the optimal forcing frequency $\mathit{\omega_{opt}}$ increased with a decrease in forcing amplitude. To explore if this effect existed for smaller amplitudes, we simulated the same system for different forcing amplitudes $\epsilon=[10^{-2}-10^{-8}]$, and plotted the breakup length $l_c$ as a function of the forcing frequency $\omega$ as shown in Fig. \ref{dns-epsilon1}, where the optimal forcing frequency for a fixed $\epsilon$, is marked with a red cross. The results show an increase in $l_c$ and $\mathit{\omega_{opt}}$ as $\epsilon$ decreases. The increase in breakup length is obvious due to the decreasing destabilizing strength of the forcing amplitude. The increase in optimal forcing frequency, however, is the most interesting observation drawn from the numerical results, since it is expected to saturate for small enough forcing amplitudes. We believe that the increase in $\mathit{\omega_{opt}}$ as $\epsilon$ decreases from $10^{-2}$ to $10^{-8}$ is a consequence of the stretched base state due to gravity, which results in the downstream stretching of the perturbation wavelength initiated at the nozzle. As the forcing amplitude decreases, the stable jet length $l_c$ increases and so does the stretching close to the jet tip. Thus to compensate for the larger stretching, the breakup potential of the forcing is sustained by increasing the forcing frequency. 

To conclude, the numerical simulations confirm the dependence of the $\mathit{\omega_{opt}}$ on the forcing amplitude, a factor generally neglected for linear stability analysis as long as $\epsilon \ll 1$. The trend also constitutes a major difference from a jet with no gravity effect ($\mathit{Bo_{in}}=0$), where $\mathit{\omega_{opt}}$ is independent of $\epsilon$ (refer Fig. \ref{dnsRes_noGravity} in Appendix \ref{app_5g}. Finally, we confirm that the preferred-mode analysis carried out for the jet is solely due to the effect of external forcing. For the given parameter range, the jet tip did not induce any self-sustained breakups.

\section{Local stability analysis}\label{LinStab}
The linear stability theory is applicable for small forcing amplitudes ($\epsilon \ll1$) and does not take into account its absolute value, a parameter that has already been shown in Section \ref{DNS_results} to influence the optimal forcing frequency. Nevertheless, we begin our analysis in the local framework by first obtaining the dispersion relation for parallel viscous jets in absence of gravity, which is used as a basis for obtaining the absolute/convective instability transition criteria in Section \ref{sec_locAna_grav}. Next, the dispersion relation for parallel jets is suitably modified to include the spatial variation of the gravitationally stretched base flow and the spatial stability of the jet is performed in Section \ref{spG1}. Since the base state is spatially evolving, we extend our stability analysis using the WKBJ formulation in Section \ref{sec_wkb}.
\subsection{Local stability analysis for jets in absence of gravity}\label{sec_locAna}
We derive the dispersion relation for the coupled equations \eqref{eq1_govEq}, governing the growth of small perturbations about the base state. Considering the normal mode expansion, the flow variables $h(z,t)$ and $u(z,t)$ are decomposed as:
\begin{subequations}
\begin{align}
h(z,t)&=h_b+\epsilon  \hat{h} e^{\mathrm{i}(kz-\omega t)},\\
u(z,t)&=u_b+\epsilon  \hat{u} e^{\mathrm{i}(kz-\omega t)},
\end{align}
\end{subequations}
where $\epsilon\ll 1$, with $\hat{h} $ and $\hat{u} $ as complex constants. $k$ and $\omega$ are respectively the dimensionless spatial wavenumber and the temporal frequency, which may both be complex. Similarly, for the variable representing the square of interface, $a(z,t)$ is decomposed as,
\begin{equation}
a(z,t)=a_b+\epsilon  \hat{a} e^{\mathrm{i}(kz-\omega t)},
\end{equation}
where $a_b=h_b^2$ and $\hat{a}=2h_b\hat{h}$. Inserting the above expansion into equation \eqref{eq1_govEq}, linearizing about $(h_b,u_b)$, and replacing $h^2\rightarrow a $, will lead to a linearised system of equations which can be formulated as an eigenvalue problem with the eigenmodes represented by $\mathbf{{\hat{q}}}=[\hat{a},\hat{u}]$.

In absence of gravity, $\mathbf{{\hat{q}}}$ represents a constant independent of $z$, $u_b=\sqrt{\mathit{We}}$ and $h_b=1$. Combining both the linearised continuity and momentum equations leads to the dispersion relation,
\begin{equation} \label{dispRel_h}
\omega = \mathit{\sqrt{We}} k - \frac{3\mathrm{i}\mathit{Oh}k^2}{2} \pm \mathrm{i} \sqrt{\frac{(k^2-k^4)}{2} + \frac{9\mathit{Oh}^2k^4}{4}},
\end{equation}
where $\mathit{Oh}$ is constant throughout the domain. The dispersion relation is used to perform a spatio-temporal stability analysis which includes the effect of advection speed of the jet on its stability properties. In this framework, we define the impulse response of a system to a localised perturbation which generates a wave packet growing in space and time. In the laboratory framework, the spatio-temporal behaviour of the wave packet can be described in terms of the complex absolute wave number $k_0$ and the corresponding complex absolute frequency $\omega_0=\omega(k_0)$ whose imaginary part $\omega_{0,i}$ will determine the temporal evolution of the wave packet. For $\omega_{0,i}>0$ the system is absolutely unstable since the disturbance grows fast enough to invade entire domain in the laboratory frame and for $\omega_{0,i}<0$ the system is convectively unstable as the localised perturbations are allowed to convect downstream before they grow in the laboratory framework. The complex pair ($k_0,\omega_0$) is defined using the saddle point condition or the Briggs-Bers zero-group velocity criterion, together with the dispersion relation 
\begin{equation}\label{saddle_pt}
\frac{d \Delta}{d k} (\omega_0,k_0) = 0, ~~~~~ \Delta(\omega_0,k_0)=0,
\end{equation}
where $\Delta$ represents the dispersion relation \eqref{dispRel_h} and 
\begin{equation}
\frac{\partial \Delta}{\partial k} = \sqrt{\mathit{We}} - 3\mathrm{i}\mathit{Oh}k \pm \mathrm{i} \frac{1-2k^2+(3\mathit{Oh}k)^2}{\sqrt{2(1-k^2)+(3\mathit{Oh}k)^2}}. 
\end{equation}
\begin{figure}
\centering
\includegraphics[trim=10 0 30 0,clip,width=0.6\linewidth]{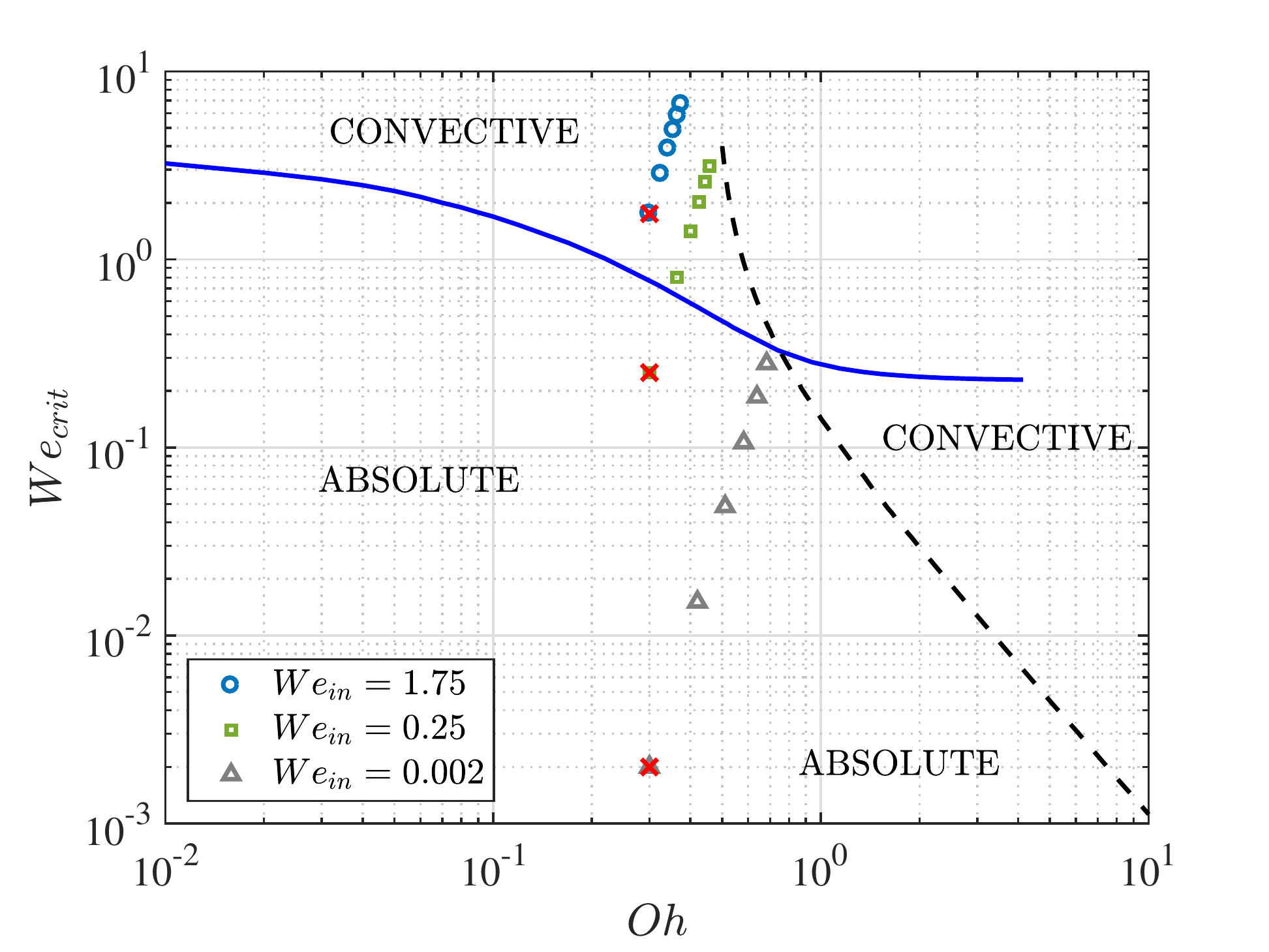}
\caption{The plot shows the absolute-convective transition (represented by full and dashed lines) for viscous jets ($\mathit{Bo_{in}}=0$). The local variation in $\mathit{Oh_z}$ and $\mathit{\mathit{We_{z}}}$ for three jets with different $\mathit{We_{in}}$ (constant $\mathit{Oh_{in}}=0.3$, $\mathit{Bo_{in}}=0.1$, $L=50$) are plotted with markers where the red cross for each represents the inlet condition at $z=0$. The distance between consecutive markers for the same jet represents an axial gap of 10 units. } 
\label{critWeCurve1}
\end{figure}%
Equation \eqref{saddle_pt} identifies the critical dimensionless speed $\mathit{We_{crit}}$, for a fixed $\mathit{Oh}$, which signifies the transition of the jet from an absolutely unstable to convectively unstable system as shown in Fig. \ref{critWeCurve1} with the full and dashed lines. The two curves are obtained for a system initialised either using a low or a high $\mathit{Oh}$, respectively. For the intermediate values of $\mathit{Oh}$ we obtain two saddle points thus giving two distinct values of the critical curve.
\subsection{Local stability analysis for jets in presence of gravity}\label{sec_locAna_grav}
\begin{figure}
\centering
\includegraphics[trim=50 0 40 0,clip,width=0.75\linewidth]{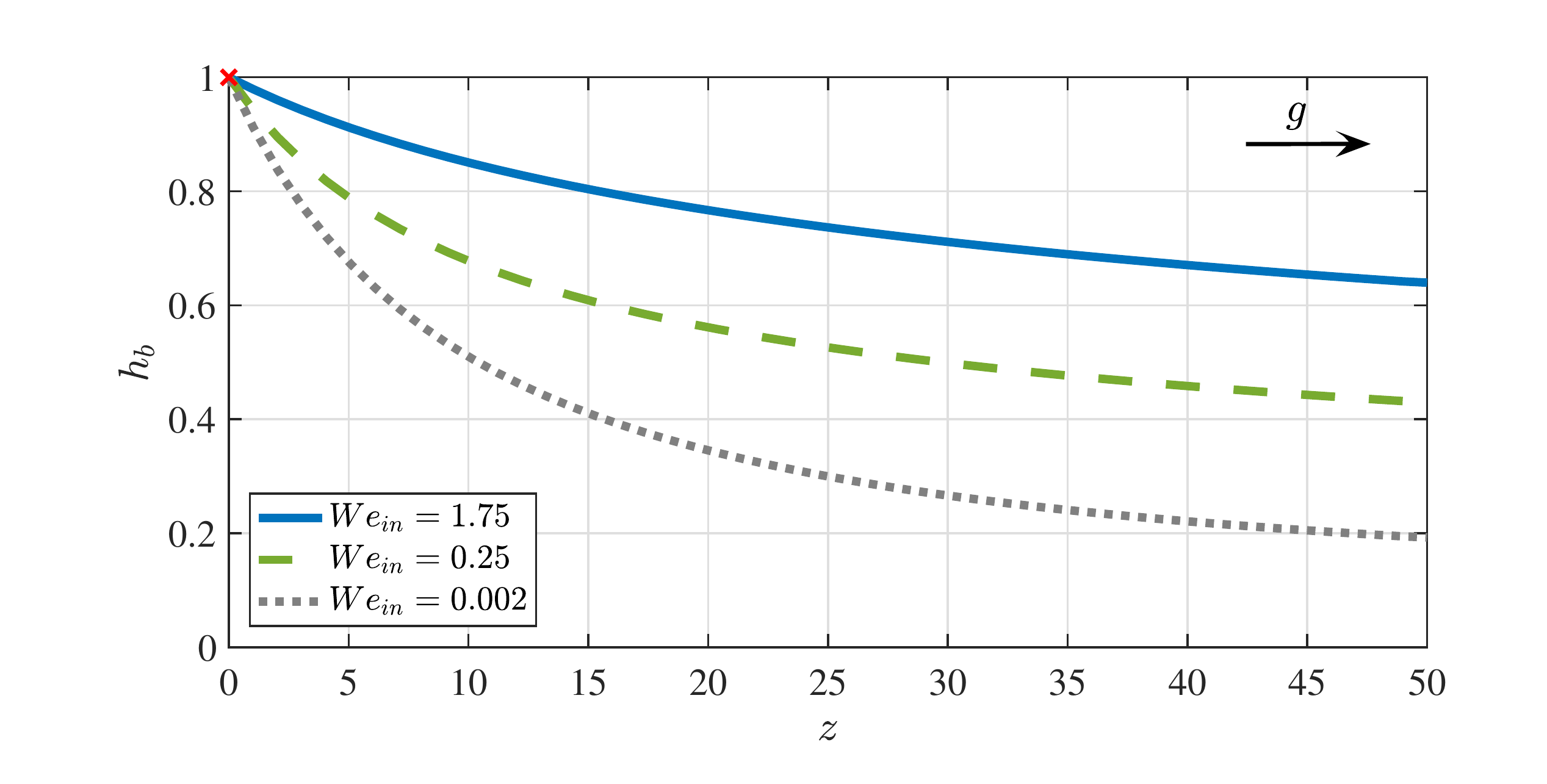}
\caption{The plot shows the stretching (or necking) close to the nozzle, of the base flow due to the presence of gravity for a jet with $\mathit{Oh_{in}}=0.3$, $\mathit{Bo_{in}}=0.1$ and three different $\mathit{We_{in}}$. Clearly, the effect of gravity is prominent for the jet with the smallest $\mathit{We_{in}}$.} 
\label{critWeCurve2}
\end{figure}

Extending the formalism for parallel jets to spatially varying jets, we derive the dispersion relation for the coupled equations \eqref{eq1_govEq}, governing the growth of small perturbations about the base state. The linearised system of equations obtained around the spatially varying base flow, can be formulated as an eigenvalue problem with the eigenmodes represented by $\mathbf{{\hat{q}(z)}}=[\hat{a}(z),\hat{u}(z)]$ as functions of $z$. Next we express the local stability of a gravity jet by introducing the terms $\mathit{Oh_z}$ and $\mathit{\mathit{We_{z}}}$, which are the local dimensionless numbers at an axial distance of $z$ from the nozzle. They are expressed as,
\begin{subequations}
\begin{align}
\mathit{Oh_{z}}&=\mathit{Oh_{in}}\sqrt{\frac{h_{b(0)}}{h_{b(z)}}},\\
\mathit{We_{z}}&=h_{b(z)}{u_b^2}_{(z)}.
\end{align}
\end{subequations}
We then plot the values $\mathit{Oh_{z}}$ and $\mathit{We_{z}}$ along the entire axial domain $L$ above the absolute-convective transition curve in Fig. \ref{critWeCurve1}. The variation in local $\mathit{Oh_z}$ and $\mathit{\mathit{We_{z}}}$ along the jet defined within a domain size $L=50$, for $\mathit{Oh_{in}}=0.3$ and $\mathit{Bo_{in}}=0.1$ and for three different inlet Weber numbers, $\mathit{We_{in}}=[1.75, 0.25,0.002]$ are represented by the markers in Fig. \ref{critWeCurve1}. The gap between consecutive markers is representative of an axial interval of 10 units. In each case, the red cross represents the inlet of the jet whose base state is shown in Fig. \ref{critWeCurve2}. We remind the reader that the case with $\mathit{We_{in}}=1.75$ corresponds to the jet whose numerical analysis has been presented in Section \ref{DNS_results}. 

For $\mathit{We_{in}}=1.75$ and $0.002$, the entire jet exists in the convective and absolute region, respectively. For intermediate $\mathit{We_{in}}=0.25$, there exists a small pocket of absolute instability close to the nozzle, after which the local parameters modify along the downstream direction resulting in the transfer of the jet into a convectively unstable regime.  

The parameter $\mathit{Bo_{in}}$ indirectly decides the instability of the jet by affecting the base state solution. Since $\mathit{Bo_{in}}$ is constant, its relative strength for the stretching of the jet interface depends on the corresponding value of $\mathit{We_{in}}$, with the effect being more pronounced for lower values of $\mathit{We_{in}}$ as shown in Fig. \ref{critWeCurve2}.

Next, for the spatially varying base flow, we perform the stability analysis in a local framework wherein the system is considered parallel at each axial location. The local dispersion relation, which now includes the local spatially varying base flow properties and is given by,
\begin{equation} \label{dispRel_h_g}
\omega^2 -2u_b(z)\omega k +\Bigg(\frac{1}{2 \sqrt{a_b(z)}}+u_b(z)^2 +3\text{i}\mathit{Oh_z}\omega\Bigg)k^2 -3\text{i}\mathit{Oh_z}u_b(z) k^3 -\frac{\sqrt{a_b(z)}}{2}k^4=0.
\end{equation}

\begin{figure}
\centering
\includegraphics[trim=40 0 40 0,clip,width=0.9\linewidth]{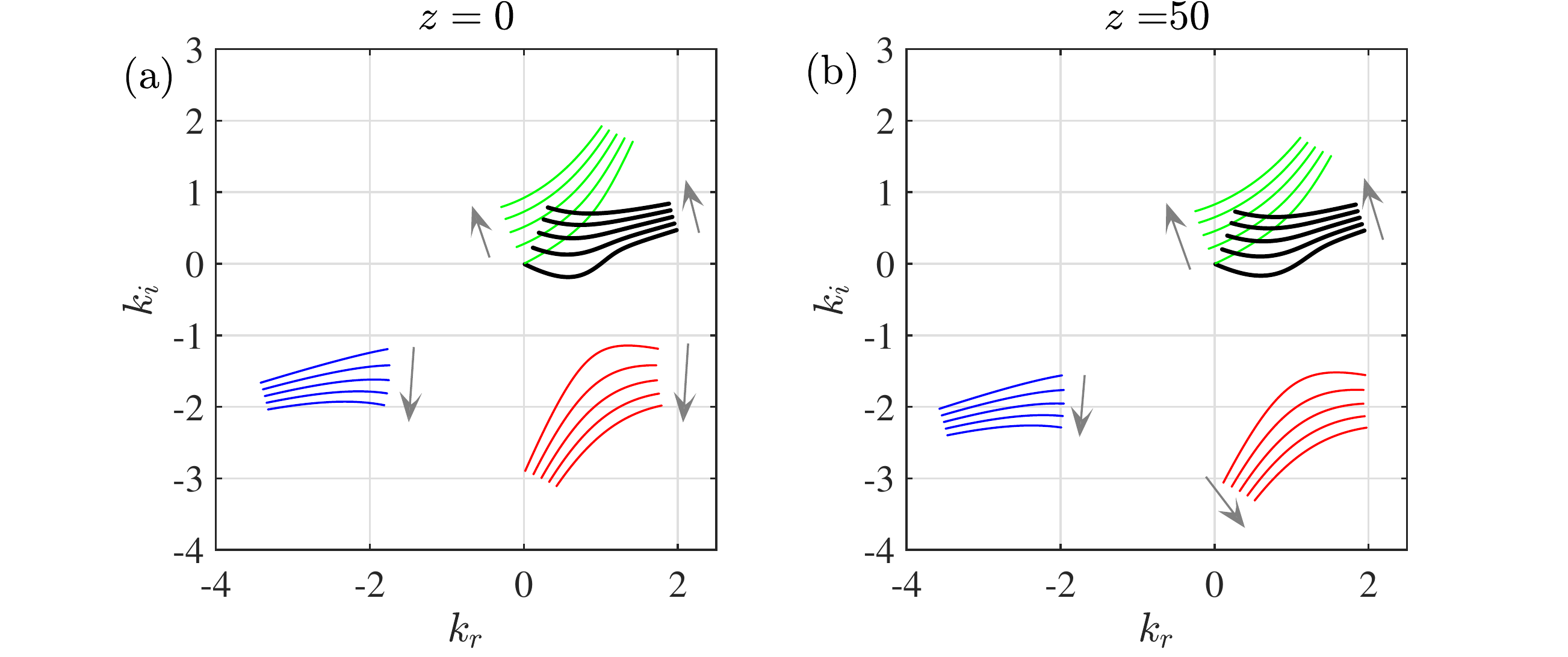}
\caption{The four $k$ branches shown in four different colours, obtained as a solution of the dispersion relation for complex $\omega$ and for increasing values of $\omega_i$ for a jet defined by $\mathit{Oh_{in}}=0.3$, $\mathit{Bo_{in}}=0.1$ and $\mathit{We_{in}}=1.75$, at (a) the nozzle outlet $z=0$ and (b) the jet exit $z=L=50$. The arrows represent the direction of movement of the waves for increasing values of $\omega_i$.} 
\label{k_waves_1}
\end{figure}

For the convectively unstable jet ($\mathit{We_{in}}=1.75$), the solution of the dispersion relation \eqref{dispRel_h_g} for a given range of complex $\omega$ (with $\omega_i>0$) results in obtaining four spatial branches which are expressed as the roots of the fourth-order polynomial \eqref{dispRel_h_g}. The solution consists of upstream (referred as $k^-$) and downstream (denoted by $k^+$) propagating branches. To identify these branches, we successively add an artificial $\omega_i$ so as to separate the branches into the upper $k_i>0$ and lower $k_i<0$ planes (\citealp{huerre1998hydrodynamic,gallaire2017}). For a downstream propagating $k^+$ branch damped in space, the associated $k_i>0$. Based on this analysis, we obtain two downstream and two upstream propagating waves for the dispersion relation \eqref{dispRel_h_g}. The $k$ branches for the localised dimensionless numbers at the nozzle inlet ($z=0$) and domain end ($z=50$) are shown in Fig. \ref{k_waves_1}(a) and \ref{k_waves_1}(b) respectively with the two $k^+$ branches denoted by the black and green colour and the two $k^-$ waves by the red and blue colour. The presence of two $k^+$ and $k^-$ waves is not specific to the present jet characteristics but rather exists for all the tested cases in the range of $\mathit{Oh_{in}}=[0.1~10]$, $\mathit{We_{in}}=[0.8~10]$, $\mathit{Bo_{in}}=[0~1]$ for $L=50$.

\subsection{Spatial stability analysis}\label{spG1}
Since the base flow with $\mathit{We_{in}}=1.75$ exists in the convectively unstable regime (see Fig. \ref{critWeCurve1}), we then proceed to analyse the base flow using the spatial stability framework, wherein the spatial growth rate for the imposed real frequency determines the flow stability. 

Given the polynomial nature of the dispersion relation, there are four spatial waves. We have verified that two of them are $k^+$, downstream propagating, waves while the remaining two are $k^-$, upstream propagating, waves (see Fig. \ref{k_waves_1}). For a more detailed account on the nature of spatial waves in capillary jets, depending on the flow model, the reader is referred to \citet{guerrero2016spatial}.

\begin{figure}
\centering
\includegraphics[trim=30 0 50 0,clip,width=1\linewidth]{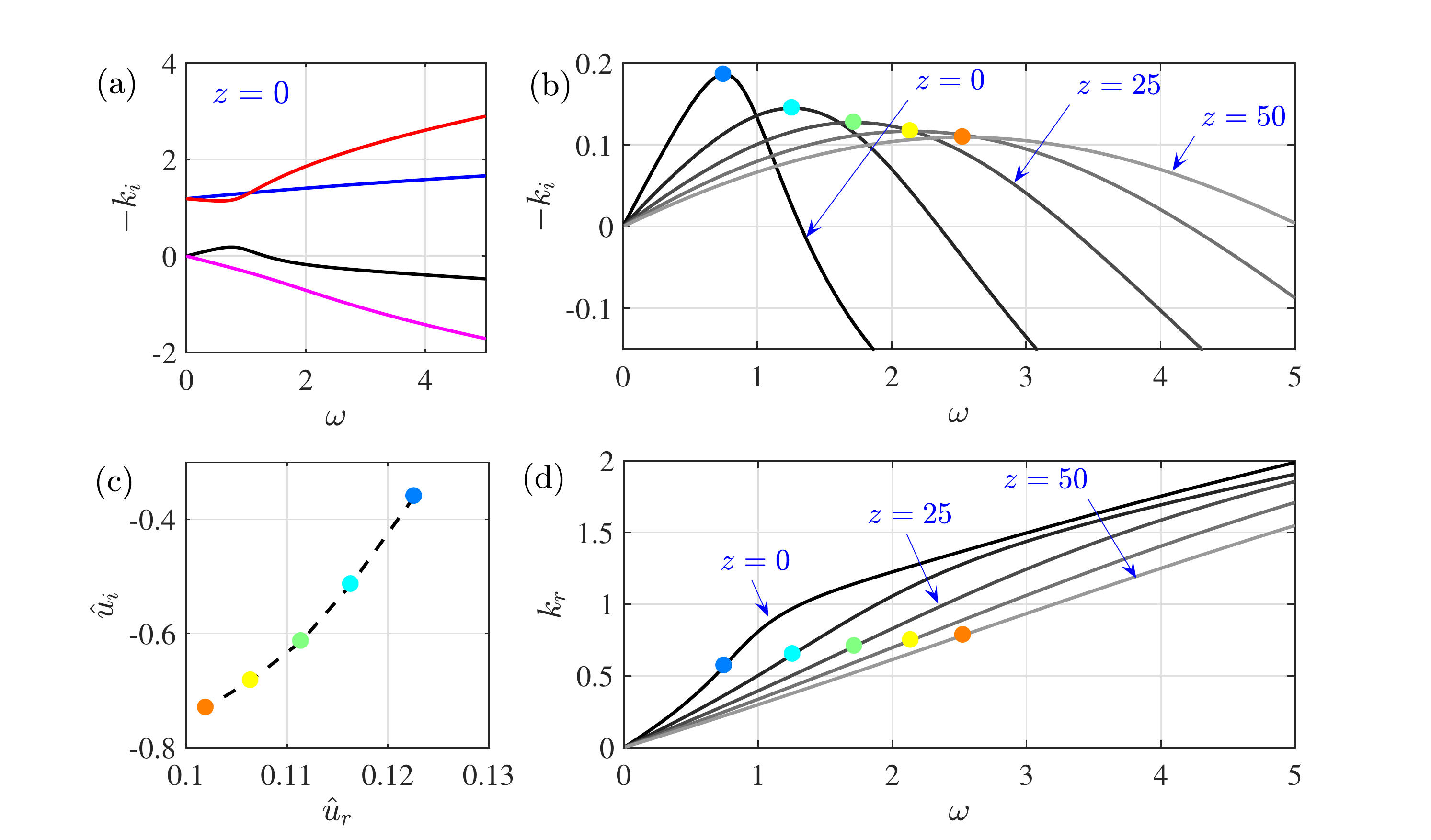}
\caption{(a) Growth rate $-k_i$ for a jet defined by $\mathit{Oh_{in}}=0.3$, $\mathit{Bo_{in}}=0.1$ and $\mathit{We_{in}}=1.75$, plotted as a function of the frequency for the four $k$ branches at the nozzle exit with the dominant $k$ branch represented in black. (b) Represents the growth rate corresponding to the dominant $k$ branch at different axial locations. (c) Shows the evolution of the locus of the real and imaginary parts of $\hat{u}_r$ and $\hat{u}_i$ at the optimal frequency as $z$ increases. (d) Represents the $k_r$ related to the optimal growth rate in (b) as a function of real frequency. The coloured markers ($\bullet$) in (b)-(d) correspond to the same forcing frequency and the axial location.} 
\label{k_waves_2}
\end{figure}

Among the four $k$ waves, only one of the $k^+$ waves is seen to be amplified. To obtain this dominant $k$ wave we plot the spatial growth rate $-k_i$ as a function of the real forcing frequency $\omega$ at nozzle inlet. As shown in Fig. \ref{k_waves_2}(a), among the four $k$ branches, only the branch denoted in black has a growth rate which is positive in its propagation direction. We chose the $k$ wave corresponding to this amplified $k^+$ branch as the dominant wavenumber for all frequencies. The relevant $k(z)$ branches are then obtained for different $z$ along the jet as shown in Fig. \ref{k_waves_2}(b) and \ref{k_waves_2}(d) by imposing the spatially dependent base flow and $\mathit{Oh_z}$ in equation \eqref{dispRel_h_g}. Fig. \ref{k_waves_2}(b) shows that the most amplified frequency shifts to higher values as one travels away from the nozzle. The associated eigenmode $\hat{\bf q}(z)$ also changes as one progresses downstream. Imposing $\norm{\mathbf{\hat{q}}}=1$ at every axial location as the normalisation condition, together with $\hat{a}_i=0$ to set the phase, we see in Fig. \ref{k_waves_2}(c) the evolution of the locus of the real and imaginary parts of the remaining degrees of freedom $\hat{u}_r$ and $\hat{u}_i$ as $z$ increases (remember that $\hat{u}_r^2+\hat{u}_i^2+\hat{a}_r^2=1$). While this locus is difficult to interpret from a physical point of view, it highlights the change of the eigenmode along the jet axis in such nonparallel gravity driven jets.

\begin{figure}
\centering
\includegraphics[trim=70 0 40 0,clip,width=1\linewidth]{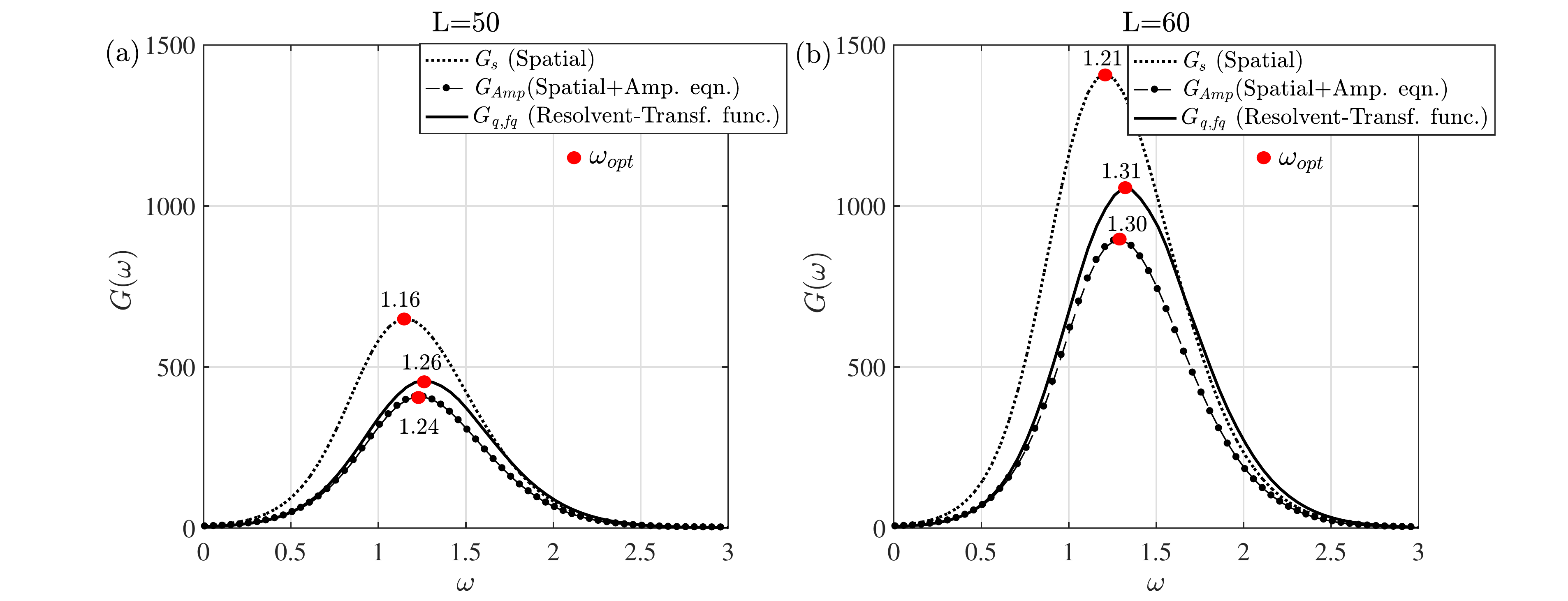}
\caption{Comparison of the total gain $G$ at different frequencies $\omega$ from the resolvent analysis and the spatial analysis for domain sizes (a)$L=50$ and (b)$L=60$ and for the jet defined by $\mathit{Oh_{in}}=0.3$, $\mathit{Bo_{in}}=0.1$ and $\mathit{We_{in}}=1.75$. The resolvent gain is computed by using the transfer function and the direct mode obtained from the spatial analysis. All the theories predict a shift in $\mathit{\omega_{opt}}$ as $L$ is increased.  } 
\label{Gain_Bo0.1}
\end{figure}

The knowledge of the relevant $k^+$ wave obtained for a given $\omega$ allows us to evaluate the leading order response due to  different forcing frequencies imposed on the base flow, conveniently expressed as
\begin{equation} 
\mathbf{q^\prime}(z,t)=\mathbf{\hat q} (\omega,z)\exp\Bigg[ \text{i}\Bigg(\int_0^z k(\omega,z^\prime)dz^\prime -\omega t\Bigg)\Bigg].
\end{equation}
The overall response norm defined in a domain size $L$ is then given as,
\begin{equation} \label{}
\mathit{G}_s(\mathit{\omega,L})= \norm {\displaystyle \int_0^L \mathbf{\hat q} (\omega,z)\exp\Bigg[ \text{i}\Bigg(\int_0^z k(\omega,z^\prime)dz^\prime -\omega t\Bigg)\Bigg]}.
\end{equation} 
This allows us to determine the optimal forcing frequency $\mathit{\omega_{opt}}$ which results in the maximal gain,
\begin{equation} \label{eq_gain_spa}
\mathit{G_{s,max}}(L)=\underset{\omega}{\text{max}}~[G_s(\omega,L)],
\end{equation} 
attained at a frequency $\omega_{opt}$. Fig. \ref{Gain_Bo0.1} (in dotted lines) shows the spatial gain as a function of forcing frequency for two arbitrary domain sizes $L=50$ and $60$. We notice that $\mathit{\omega_{opt}}$ shifts from 1.16 to 1.21 as we increase the domain size. 
\subsection{Weakly nonparallel stability analysis (WKBJ)}\label{sec_wkb}
In order to further incorporate the non-parallelism of the base flow, we extend our spatial analysis by including the WKBJ formalism introduced by \citet{gaster1985large} and \citet{huerre1998hydrodynamic} for a spatial mixing layer and applied by \citet{viola2016mode} for swirling flows. 

In this framework, we introduce a slow streamwise scale $Z$, which relates to the fast scale $z$ as $Z=\eta z$, where $\eta\ll1$ is a measure of the weak non-parallelism. The new base flow depends only on $Z$ and the global response to inlet forcing takes the modulated wave form:
\begin{equation}
\mathbf{q^\prime}(Z,t)\sim A(Z)\mathbf{\hat q} (Z,0)\exp\Bigg[ \text{i}\Bigg(\frac{1}{\eta}\int_0^Z k(\omega,Z^\prime)dZ^\prime -\omega t\Bigg)\Bigg],
\end{equation}
where $\mathbf{\hat q}(\omega,Z)$ is the local eigenmode and $k(\omega,Z)$ the local wavenumber at section $Z$ and a fixed forcing frequency $\omega$. The amplitude function $A(Z)$ acts as an envelope, smoothly connecting the progressive slices of the parallel spatial analysis. At each axial location, we impose $\mathbf{\hat q}^{H} \cdot \mathbf{\hat q}=1$, where $(\cdot)^H$ is the transconjugate. As described in Appendix \ref{wkbj_all}, imposing an asymptotic expansion and a compatibility condition, the local stability analysis is retrieved at zeroth-order in $\eta$, while at first order in $\eta$ the following amplitude equation is obtained:
\begin{equation}\label{eq_amp}
M(Z)\frac{\mathrm{d}A(Z)}{\mathrm{d}Z} +N(Z) A(Z) = 0,
\end{equation}
 whose solution is given as 
 \begin{equation}\label{eq_amp_sol}
A(Z)=A_0 \exp\Bigg( -\int_0^Z \frac{N(Z^\prime)}{M(Z^\prime)} \mathrm{d}Z^\prime \Bigg).
\end{equation}
The functions $M(Z)$ and $N(Z)$ are defined in Appendix \ref{wkbj_all}. The amplitude at the inlet is set as, $A(0)=1$ which simplifies the forcing expression at the inlet to $\mathbf{q}^\prime(0,t)= \mathbf{\hat{q}}(0)\exp(i\omega t)$. Finally we express the total spatial gain at first order as
\begin{equation} \label{wkbj_gain}
\mathit{G}_{Amp}^2(\mathit{\omega,L})=\frac{\displaystyle \int_0^z A^H(z^\prime)A(z^\prime)\Big( \mathbf{\hat q}^{H}(z^\prime) \cdot \mathbf{\hat q} (z^\prime)\Big) \Big(e^{\int_0^{z^\prime} -2k_i(z^{\prime \prime}) \mathrm{d}z^{\prime \prime}}\Big) \mathrm{d} z^\prime}{\mathbf{\hat q}^{H}(0) \cdot \mathbf{\hat q}(0)}.
\end{equation} 
The global gain of the response due to the forcing frequency, for fixed domain sizes, is reported in Fig. \ref{Gain_Bo0.1}, where $\mathit{\omega_{opt}}= 1.24$ and $1.30$ for $L=50$ and $60$, respectively. The WKBJ approximation greatly modifies the gain when compared to the zeroth-order analysis and shifts the optimal forcing frequency predicted from the spatial analysis which excludes the amplitude equation. However, to truly assess the validity of the amplitude equation one needs to analyse the base flow in the global framework using the resolvent analysis, which will be the focus of our discussion in the next section.

\section{Global stability analysis}\label{global}
Unlike the local stability analysis, the global stability framework allows taking into consideration the axially varying base state due to the stretching effect of gravity. In this framework, we first evaluate the inherent global stability of the base flow in Section \ref{global_stability}. We next perform a resolvent analysis in Section \ref{resol_ana} on the globally stable base flow to evaluate its response in presence of a given perturbation. 
\subsection{Global stability}\label{global_stability}
Since the base flow is spatially varying, the perturbations imposed on it are no longer sought in the form of Fourier modes but are expanded in the form:
\begin{subequations}\label{global_mode}
\begin{align}
h(z,t) &= h_b(z) + \epsilon \tilde{h}(z)e^{\lambda t},\\
u(z,t) &= u_b(z) + \epsilon \tilde{u}(z)e^{\lambda t},
\end{align}
\end{subequations}
where $\epsilon \ll 1$ and $\tilde{h}(z),\tilde{u}(z)$ are the global stability modes related to the complex growth-rate $\lambda$. Substituting expressions \eqref{global_mode} in equations \eqref{eq1_govEq} and linearising around the base state $(h_b,u_b)$ results in the general eigenvalue problem of the form
\begin{equation}\label{eq4}
   \lambda {\rm I\!I}\begin{bmatrix}\tilde{h} \\
                                          \tilde{u}
                 \end{bmatrix}
                  =  
  \mathit{M} \begin{bmatrix} \tilde{h}\\
                                                \tilde{u}
                       \end{bmatrix},
\end{equation}
with boundary conditions $\tilde{h}(0,t)=0$ and $\tilde{u}(0,t)=0$. We do not impose any boundary conditions at the end of domain $z=L$ since it is not possible \textit{a priori} to distinguish between amplifying perturbations and transient disturbances. This will occur in any problem that involves an `active system' and can support amplifying waves (\cite{briggs1964electron} and \cite{leib1986generation}). The global stability analysis presented in \cite{Rubio} does not impose any boundary conditions for $z=L$ since the numerical method naturally converges to the most regular asymptotic solution of the base flow equation \eqref{eq2_finalBS} and eigenvalue problem \eqref{eq4}, as $z\to \infty$. Nonetheless, we checked that the dominant eigenvalue and eigenmode were unaffected by the presence of Neumann boundary condition at $z=L$, namely $\frac{d\tilde{u}}{dz}(L)=0$.

The complete expressions for the linear operator $\mathit{M}$ can be found in Appendix \ref{app_6.1}. The solution of equation \eqref{eq4} results into a set of eigenmodes ($\tilde{h},\tilde{u}$), whose growth rate and frequency are given by the real ($\lambda_r$) and imaginary ($\lambda_i$) parts of the related eigenvalue. A base state is stable to self induced oscillations provided $\lambda_r<0$. 

To solve the eigenvalue problem, the Chebyshev collocation method is used for obtaining the differential operators. Derivatives with respect to z are calculated using the standard Chebyshev differentiation matrices. Denoting the non dimensional physical domain as $L$, the domain is mapped into the interval $-1\leqslant y \leqslant 1$ by using the transformation $z=[(L/2)\times(y+1)]$. A validation of the global scheme with the results of \cite{Rubio} is presented in Appendix \ref{app_6.2}.  
\begin{figure}
\centering
\includegraphics[trim=50 10 60 0,clip,width=1\linewidth]{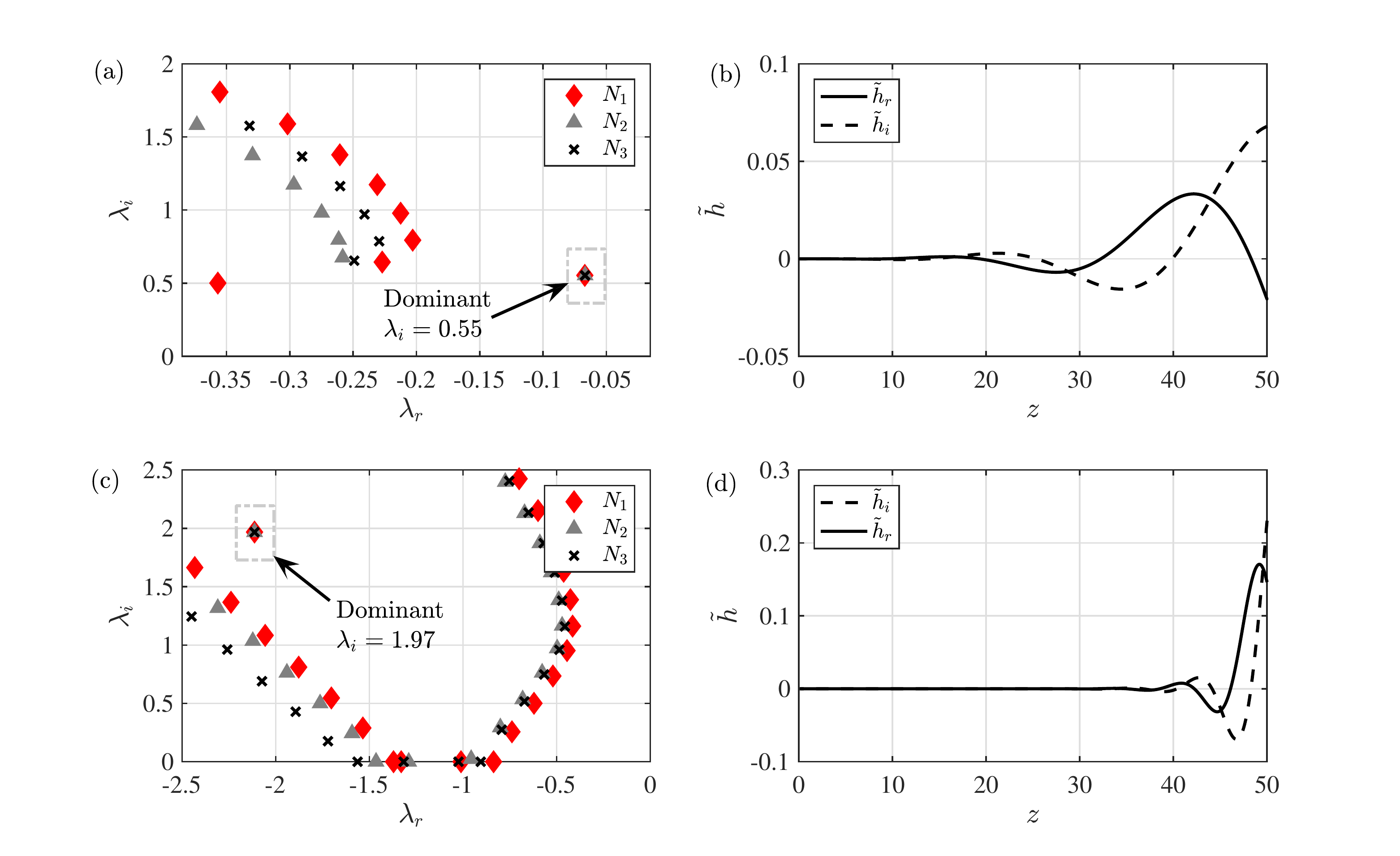}
\caption{(a),(c)Eigenvalue spectrum $\lambda$ obtained for three different nodes $N_1=100$, $N_2=125$ and $N_3=150$ and (b),(d)the real and imaginary parts of the leading eigenfunction $\tilde{h}$, for $\mathit{Oh_{in}}=0.3$, $\mathit{Bo_{in}}=0.1$, $L=50$ and evaluated for two different values of inlet Weber. (a),(b) Corresponds to $\mathit{We_{in}}=0.25$ where the leading eigenvalue has an eigenfrequency $\lambda_i=0.55$ (c),(d) Corresponds to $\mathit{We_{in}}=1.75$ with $\lambda_i=1.97$. For both the Weber numbers, the entire spectrum has a $\lambda_r<0$ rendering the system to be globally stable.} 
\label{global_stable}
\end{figure}%

\begin{figure}
\centering
\includegraphics[trim=60 0 60 0,clip,width=1\linewidth]{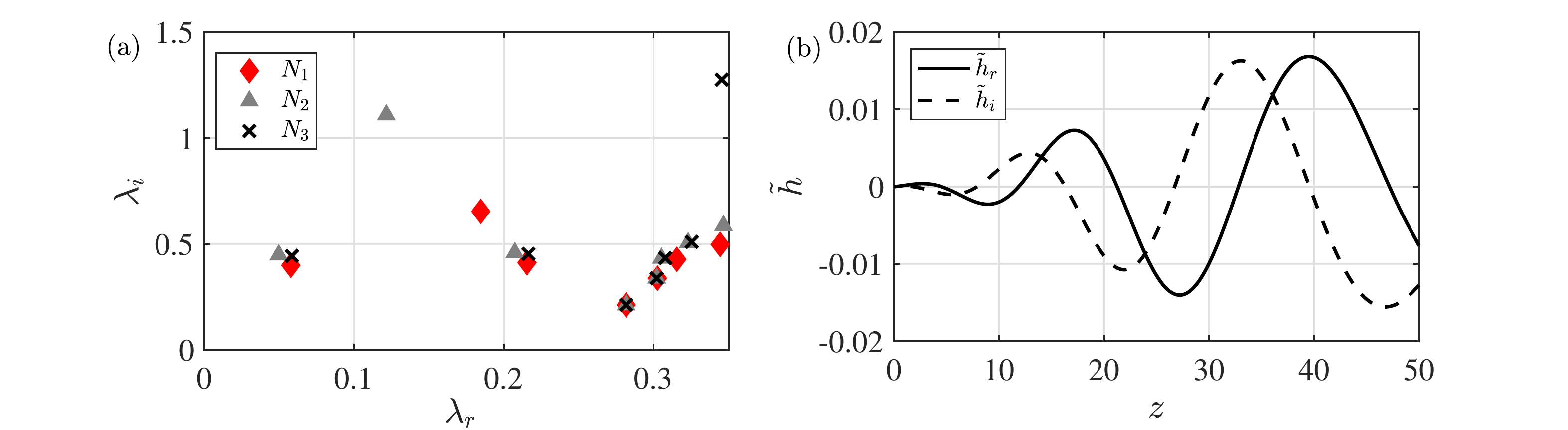}
\caption{(a) Eigenvalue spectrum $\lambda$ obtained for three different nodes $N_1=100$, $N_2=125$ and $N_3=150$ and (b) the real and imaginary parts of the leading eigenfunction $\bar{h}$, for $\mathit{Oh_{in}}=0.3$, $\mathit{Bo_{in}}=0.1$, $\mathit{We_{in}}=0.002$ and $L=50$. We note that the leading eigenvalue has a positive growth rate $\lambda_r>0$ thus rendering the system to be globally unstable.} 
\label{global_unstable}
\end{figure}

For the three cases of jets described in Fig. \ref{critWeCurve2}, with $L=50$, $\mathit{Oh_{in}}=0.3$, $\mathit{Bo_{in}}=0.1$ and three different values of $\mathit{We_{in}}$ a global stability analysis is carried out using different resolutions ($N_1=100,N_2=125,N_3=150$) to exclude spurious eigenvalues. The eigenvalue spectrum for $\mathit{We_{in}}=0.25,1.75 ~\text{and}~ 0.002$, are represented in Fig. \ref{global_stable}(a), \ref{global_stable}(c) and \ref{global_unstable} respectively. The dominant eigenvalues have $\lambda_r<0$ (for $\mathit{We_{in}}=0.25$ and $1.75$) and $\lambda_r>0$ (for $\mathit{We_{in}}=0.0025$) thus representing globally stable and unstable jets, respectively. Note however that the local stability analysis of the globally stable flow with $\mathit{We_{in}}=0.25$ predicts the jet to have a small `pocket' of absolute instability close to the nozzle. 

Eigenmodes corresponding to the dominant eigenvalues are presented in the accompanying figure. We note that the dominant eigenmode, as represented in \ref{global_stable}(b), \ref{global_stable}(d) and \ref{global_unstable}, has an amplitude that grows downstream. Fig. \ref{global_unstable}(b) also shows that the wavelength grows downstream. This is a consequence of the fluid acceleration caused by gravity (\citealp{tomotika1936breaking} and \citealp{Rubio}) and can be interpreted from Fig. \ref{k_waves_2}(d) where $k_r$ is seen to decrease with increasing $z$ for a fixed forcing frequency $\omega$. Further we see that close to the outlet, the eigenmodes evolve at a much larger length scale compared to that related to the variations in steady state jet, thus strengthening the argument that weakly non parallel stability analysis should be used with care for predicting the global stability of the gravity jet.
\subsection{Global resolvent}\label{resol_ana}
Analysing the linear response of the base state for an external harmonic forcing at frequency $\omega$ is only well defined if the linear operator is stable or in other words the base state is stable, where the imposed perturbations are allowed to travel downstream before spreading in the entire domain under consideration. Else the algebraically amplified solution is superimposed by the unforced naturally growing exponential mode. Keeping this in mind, in this section we present the resolvent analysis for the stable gravity jet ($\mathit{We_{in}}=1.75$). To compare our results with the nonlinear simulations of Section \ref{chap3_dns}, we impose similar inlet forcing conditions and approximate the gain predicted by the resolvent analysis in terms of the forcing amplitude. 
\subsubsection{Problem formulation}
The external force $f$ is modelled as an incoming perturbation in the form of an unsteady upstream boundary condition of the 1D Eggers \& Dupont equation \eqref{eq1_govEq}. The resulting linearised equation is represented as,
\begin{equation}\label{eq_res1}
\rm I\!I\partial_t{[\vect{s}]}=\mathit{M}[\vect{s}] +\mathit{{B_f}f},
\end{equation}
where as in the eigenvalue problem \eqref{eq4} $\vect{s}=[h,u]$, $\rm I\!I$ represents the identity matrix, $\mathit{M}$ the linear operator (detailed in Appendix \ref{app_6.1}) and $\mathit{B_f}$ the operator which expresses the effect of the inlet forcing onto the bulk equation. Considering a time-harmonic forcing, $f=\tilde{f} \text{exp}(-\text{i}\omega t)$, results in an asymptotic flow response $\vect{s}=\tilde{\vect{s}} \text{exp}(-\text{i}\omega t)$ at the same frequency. Here $\tilde{\vect{s}}=[\tilde{h},\tilde{u}]$. Imposing these transformations in \eqref{eq_res1} we obtain,
\begin{equation}\label{eq_res2}
-(\mathit{M}+\text{i}\omega \rm I\!I)\tilde{\vect{s}}=\mathit{{B_f} \tilde{f}}.
\end{equation}
Equation \eqref{eq_res2} is subjected to two inlet and two outlet boundary conditions. The solution forced only in $u$ at the nozzle satisfies $\tilde{h}=0$ and $\tilde{u}=1$, while for the one that is forced in both $h$ and $u$, $\tilde{h}$ and $\tilde{u}$ can be chosen arbitrarily. We use the former when comparing the results with the nonlinear simulations (where the forcing was applied using the form \eqref{eq3_dnsForcing}) whereas the latter when comparing to the spatial and WKBJ analysis of Section \ref{spG1} and \ref{sec_wkb}. 

Based on the results of the local and global analysis at $\mathit{We_{in}}=1.75$, the convective instability of the flow ensures that at the outlet any existing $k^+$ branch, obtained from the spatial analysis, will be transmitted downstream. Since the relevant $k^+(\omega,L)$ branch for a given $\omega$ and at $z=L$ can be obtained from the local analysis of Section \ref{spG1}, we impose for the solution of the equation \eqref{eq_res2} at $z=L$ the spatial response, specifically, $\tilde{h}(L)=\hat{h}\exp(\text{i}kL)$ and $\tilde{u}(L)=\hat{u}\exp(\text{i}kL)$, $k$ being the unique root of the dispersion relation corresponding to a donwstream amplified wavenumber. It should be noted that for active systems, such as the jet falling under the influence of gravity, it is not possible \textit{a priori} to impose unique boundary conditions at the outlet. Thus, substitution of the resolvent response by the spatial response at the outlet should be treated as an approach to close the differential problem of equation \eqref{eq_res2} rather than depicting the physical boundary conditions. To ensure that these boundary conditions do not affect the final response over a given domain size $L$, we impose them for a domain size $L^{\prime}>L$, such that the response for all the frequencies over $L$ is independent of the imposed boundary condition. 

Finally we express the magnitude of the response $\tilde{\vect{s}}$ due to the externally applied forcing in terms of the gain $G$, with the maximum gain expressed as,
\begin{equation}\label{eq_gain_1}
G^2_{\mathit{max}}(L)=\underset{\omega}{\text{max}}\frac{\norm{\tilde{\vect{s}}}^2}{\lVert{\tilde{f}\rVert}^2}=\underset{\omega}{\text{max}} \frac{\norm{(\mathit{M}+\text{i}\omega \rm I\!I)^{-1}\mathit{B_f \tilde{f}}}^2}{\lVert{\tilde{f}\rVert}^2},
\end{equation}
attained at $\omega_{opt}$. To measure the amplitude of the response and the forcing, we define $\mathit{Q}$ and $\mathit{Q}_f$ as the weight matrices of the discretised energy norm ($\lVert{\tilde{\vect{s}}\rVert}^2=\vect{s}^\dagger \mathit{Q} \vect{s}$) and the forcing norm ($\lVert{\tilde{f}\rVert}^2=f^\dagger \mathit{Q}_f f$), respectively, obtained for the Chebyshev space on the physical domain $L$ which is mapped into the interval $-1\leqslant y \leqslant 1$ by using the transformation $z=[(L/2)\times(y+1)]$. $\mathit{Q_f}$ is a $2N \times 2N$ matrix enabling us to distinguish forcing on $u$ only or on both components $u$ and $h$.  Following the optimization method using singular value decomposition (SVD) described in \citet{Marquet10} and \cite{Garnaud13}, we then express the optimal gain using the following eigenvalue problem,
\begin{equation}\label{eig_garnaud}
\mathit{Q}^{-1}_f\mathit{B}^\dagger_f({M}+\text{i}\omega \rm I\!I)^{-1\dagger}\mathit{Q}^\dagger(\mathit{M}+\text{i}\omega\rm I\!I)^{-1}\mathit{B_f \tilde{f}}=\lambda \mathit{\tilde{f}},
\end{equation}
whose leading eigenvalue solution $\lambda$ gives $G^2_{\mathit{max}}$ and the associated eigenmode solution $\tilde{f}$ yields the optimal normalised forcing amplitude in $(h,u)$ to be applied at the inlet. 

Since the linear analysis is based on small perturbations, the exact amplitude of the perturbation is unaccounted for in the expression \eqref{eq_gain_1}. For the resolvent analysis we then define $\mathit{G_{h,fu}}$ as the gain in ${h}$ from a solution forced only in $u$ and $\mathit{G_{q,fq}}$ as the gain in $\vect{{q}}$ for a solution forced in $\vect{{q}}$, where $\vect{{q}}=[{a},{u}]$. The two expressions for the gain: $\mathit{G_{h,fu}}$ and $\mathit{G_{q,fq}}$ are formulated to replicate the forcing and gain definitions in the nonlinear simulations (Section \ref{chap3_dns}) and the spatial analysis (Section \ref{spG1} and \ref{sec_wkb}), respectively. 

\begin{figure}
\centering
\includegraphics[trim=60 0 80 0,clip,width=1\linewidth]{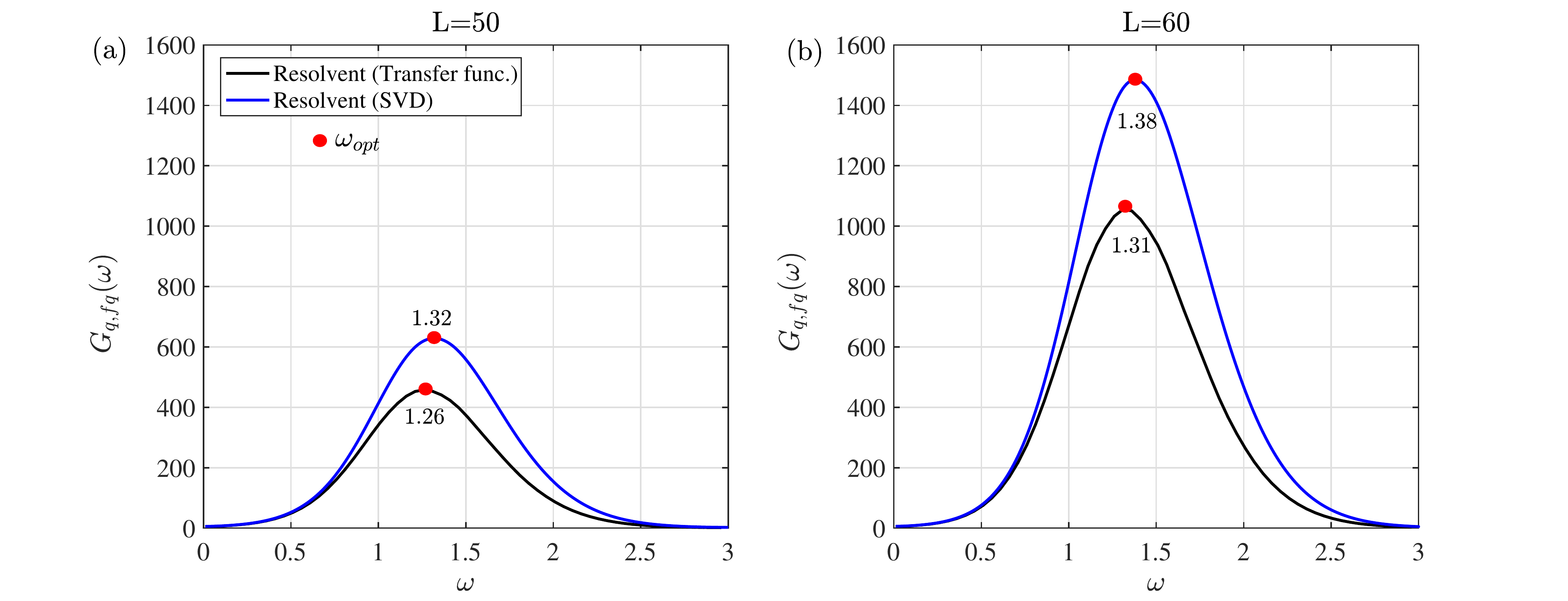}
\caption{Comparison of the total gain at different frequencies from the resolvent analysis for domain sizes (a) $L=50$ and (b) $L=60$ and for the jet defined by $\mathit{Oh_{in}}=0.3$, $\mathit{Bo_{in}}=0.1$ and $\mathit{We_{in}}=1.75$. The gain computed by applying the transfer function on the direct eigenmode from the spatial analysis is shown in black and the maximal optimal gain computed through the singular value decomposition analysis is shown in blue.} 
\label{Gain_svd_tf}
\end{figure}

Looking for the gain in $\mathit{G_{q,fq}}$ requires the inclusion of additional operators $\mathit{P}$ and $\mathit{H}$ which express $\vect{{q}}$ in terms of $\vect{{s}}$, and are given by,
\begin{subequations}\label{newOp}
\begin{align}
\vect{\tilde{q}} &= \mathit{P}\vect{\tilde{s}},\\
\tilde{f}_q &= \mathit{H}\tilde{f}.
\end{align}
\end{subequations}
The operator $\mathit{H}$ modifies the imposed boundary conditions in terms of $\tilde{a}$, whereas the operator $\mathit{P}$ adequately expresses the response in $\tilde{a}$ in terms of $\tilde{h}$ such that $\tilde{a}=2h_b\tilde{h}$. Additionally, to have an explicit comparison with the spatial analysis, we apply a forcing at inlet which is obtained as the eigenmode solution of the spatial problem in Section \ref{spG1}. Thus,
\begin{equation}
\tilde{f}_q=\mathbf{\hat q} (z=0).
\end{equation}
The gain for the imposed forcing is then obtained as,
\begin{equation}\label{eq_gain_q}
G^2_{\mathit{q,fq}}(\mathit{\omega})=\frac{\norm{\vect{\tilde{q}}}^2}{\lVert{\tilde{f_q}\rVert}^2}= \frac{\norm{\mathit{P}(\mathit{M}+\text{i}\omega \rm I\!I)^{-1}\mathit{B_f}  \mathit{H}^{-1}\mathit{\tilde{f}_q}}^2}{\lVert \tilde{f}_q \rVert^2}.
\end{equation}
Note however that even though this formalism allows a direct comparison with the spatial analysis, the gain $\mathit{G_{q,fq}}(\mathit{\omega})$ does not represent the maximum optimal gain since we do not impose the optimisation of the inlet forcing vector using an SVD formalism as was done in \eqref{eig_garnaud}. Fig. \ref{Gain_svd_tf} demonstrates the difference between the gain and $\mathit{\omega_{opt}}$ computed using the direct mode from spatial analysis (in black) and through an optimisation problem which solves for the optimal mode (in blue). Indeed the resolvent gain based on the optimised mode is much larger in magnitude.
\subsubsection{Results: Comparison with spatial stability analysis}
\begin{figure}
\centering
\includegraphics[trim=40 0 40 0,clip,width=0.85\linewidth]{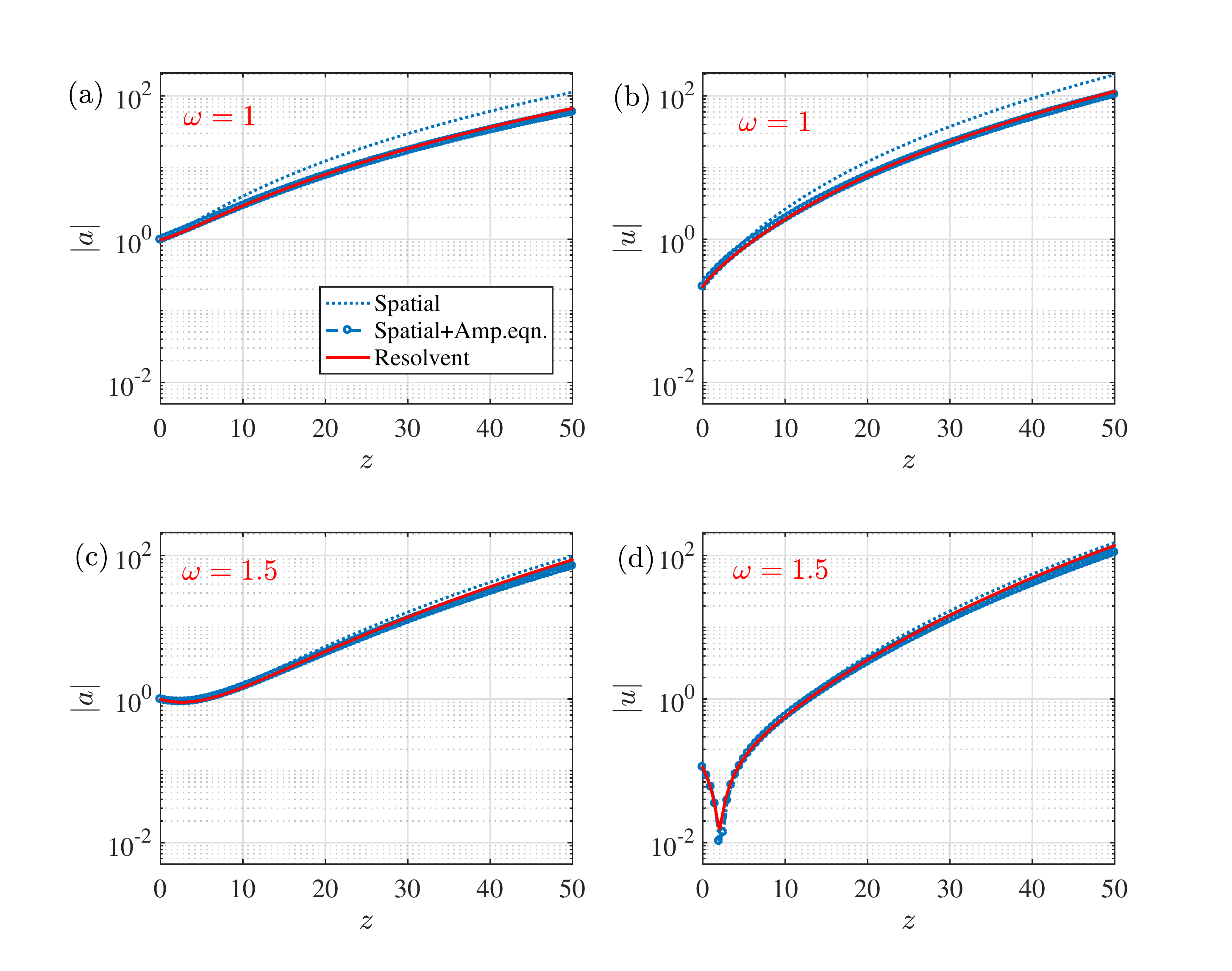}
\caption{Resolvent and spatial response ($|a|$ and $|u|$) of the jet characterised by $\mathit{Oh_{in}}=0.3$, $\mathit{We_{in}}=1.75$ and $\mathit{Bo_{in}}=0.1$ at (a)-(b) $\omega=1$ and (c)-(d) $\omega=1.5$ with a domain size $L=50$.}
\label{g_resp_1.3}
\end{figure}

To replicate the type of forcing and the expression of gain used in the spatial analysis in Section \ref{spG1} and \ref{sec_wkb}, we impose the eigenmode solution at the nozzle exit obtained from the spatial analysis as the forcing vector in the resolvent analysis. The resulting gain $\mathit{G_{q,fq}}$ for two different fixed domain sizes $L=50$ and $L=60$ are shown in Fig. \ref{Gain_Bo0.1}. We observe from the figure that the inclusion of the amplitude equation in evaluating the spatial response by far improves the estimation of the true gain obtained from the resolvent analysis. Moreover the predicted $\mathit{\omega_{opt}}$ producing the largest $\mathit{G_{q,fq}}$ from the WKBJ analysis is in close agreement with that of the resolvent analysis. The response norm obtained using the different approaches agrees qualitatively (see Fig. \ref{g_resp_1.3}). Its non-monotonic behaviour at $\omega=1.5$ (see Fig. \ref{g_resp_1.3}(c-d)) is well captured and is in accordance with the work presented in \citet{LizziPhD}. However, the difference in gain between the three methods originates as a result of the quantitative disparity in the response obtained at different frequencies. The divergence between the spatial and resolvent analysis is due to the stretching effect of gravity on the base flow. For a parallel base flow, the results obtained from both the methods are found to be identical (as shown in Appendix \ref{app_5g}, Fig. \ref{Spat_gain}).
\subsubsection{Results: Comparison with nonlinear simulations}
We compare the resolvent analysis with the nonlinear simulations of Section \ref{DNS_results}. Classically, the optimal forcing frequency $\mathit{\omega_{opt}}$ resulting in the maximum gain, can be deduced by plotting the gain $\mathit{G_{h,fu}}$ as a function of $\omega$ for a fixed domain size $L$. For capillary jets however, the domain size over which the perturbation grows cannot be fixed \textit{a priori}. It is merely an outcome of the analysis which should compare well with the value of $l_c$ measured in the nonlinear simulations. 

\begin{figure}
\centering
\includegraphics[trim=50 20 60 40,clip,width=1\linewidth]{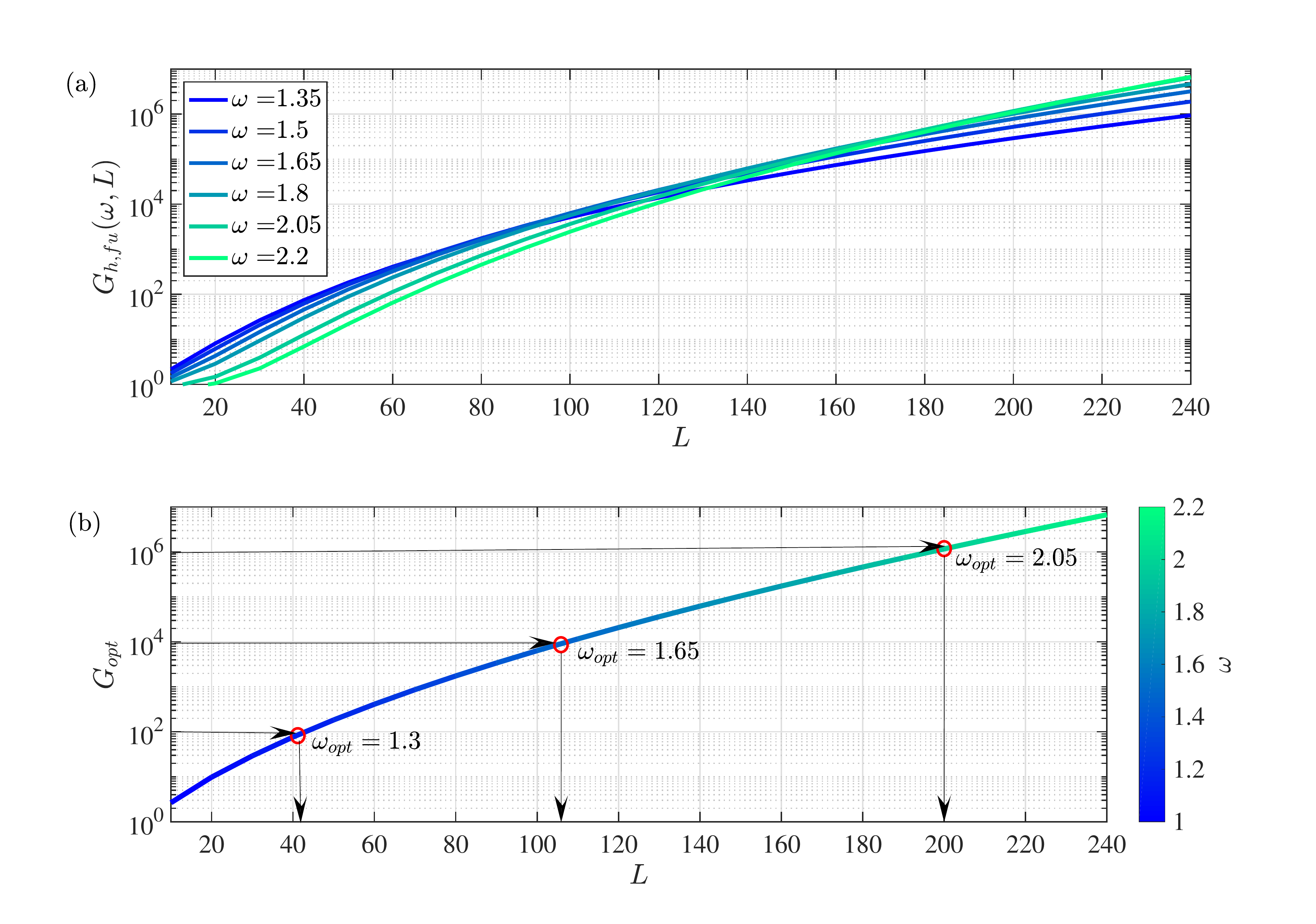}
\caption{(a) Resolvent gain computed for $\tilde{h}$ with a forcing applied only in $u$ for different values of domain sizes for a jet defined by $\mathit{Oh_{in}}=0.3$, $\mathit{Bo_{in}}=0.1$ and $\mathit{We_{in}}=1.75$. Each curve is representative of a constant frequency. (b) The dominant frequency envelope as a function of the domain size $L$. The gain represented by $10^2, 10^4$ and $10^6$ is related to forcing amplitudes $\epsilon=10^{-2}, 10^{-4}$ and $10^{-6}$ respectively. A horizontal projection from the respective gain on the frequency envelope yields the $\mathit{\omega_{opt}}$ and a vertical projection from the $\mathit{\omega_{opt}}$ on $L$ determines the breakup length $l_c$.}
\label{ResolventEps1}
\end{figure}

In order to circumvent this lack of consistency and in absence of the knowledge of $l_c$, we first plot $\mathit{G_{h,fu}}$ as a function of increasing domain sizes $L$ and for fixed $\omega$ as shown in Fig. \ref{ResolventEps1}(a) where the gain $\mathit{G_{h,fu}}(\omega,L)$ is computed for $\omega=[1-2.5]$ with $\Delta \omega=0.01$ and for $L=[10-240]$ with $\Delta L=10$. This results in a bundle of constant frequency curves, intersecting each other at different locations in $L$. In Fig. \ref{ResolventEps1}(a) we now define the dominant frequency at a given $L$ as the frequency with the maximum gain at $L$. A close examination reveals that there is a continuous transition in the dominant frequency as one moves along increasing domain sizes. This is shown in Fig. \ref{ResolventEps1}(b) where for clarity we plot only the envelope $\mathit{G_{opt}}(L)$ of the dominant frequency for all values of $L$. $\mathit{G_{opt}}(L)$ is attained for $\mathit{\omega_{opt}}(L)$. 

\begin{figure}
\centering
\includegraphics[trim=45 0 60 0,clip,width=0.9\linewidth]{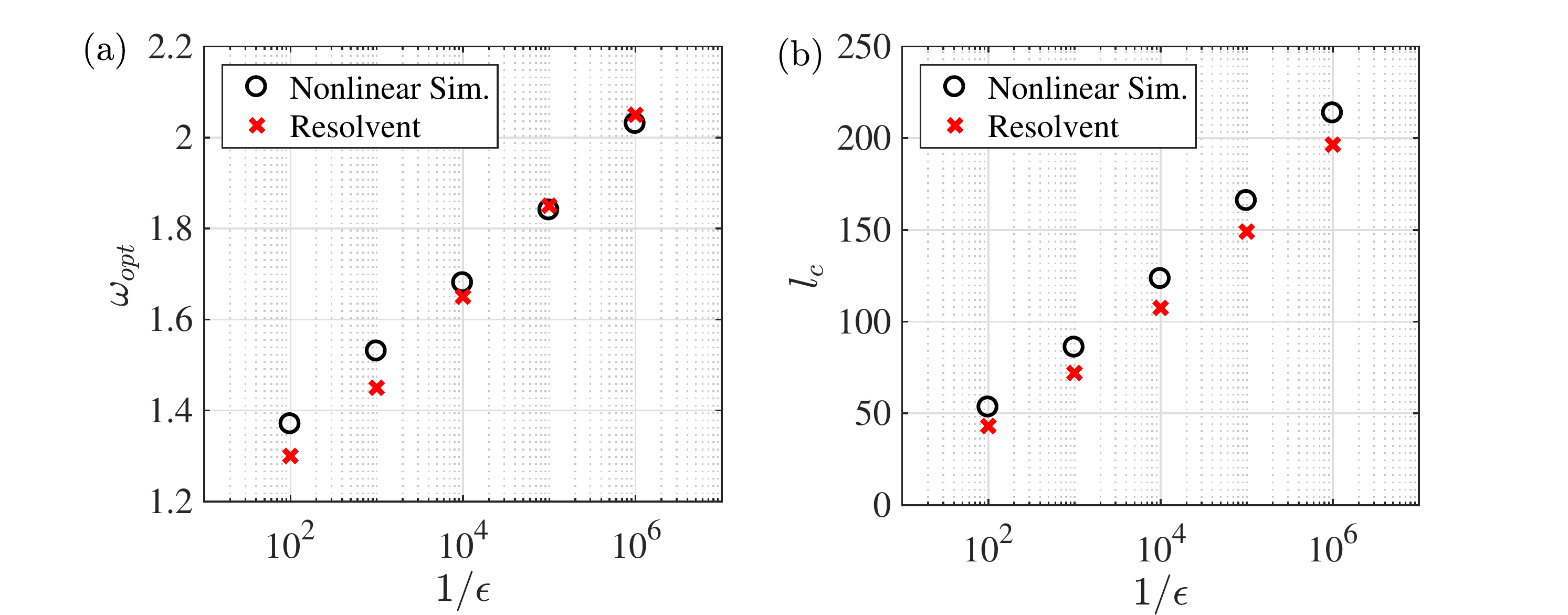}
\caption{Comparison of breakup characteristics obtained from the nonlinear simulations (Fig. \ref{dns-epsilon1}) and the resolvent analysis (Fig. \ref{ResolventEps1}) for a jet defined by $\mathit{Oh_{in}}=0.3$, $\mathit{Bo_{in}}=0.1$ and $\mathit{We_{in}}=1.75$ for (a) the optimal forcing frequency $\mathit{\omega_{opt}}$ and (b) the breakup length $l_c$ at different inverse forcing amplitudes $1/\epsilon$. }
\label{ResolventEps2}
\end{figure}

As discussed previously, $L$ represents the breakup location along the jet where the nonlinear effects appear. Broadly speaking, nonlinearity enters the system when a small perturbation $\epsilon$ gives rise to a response of the order of 1, which suggests to approximate $l_c$ by the value of $L$ at which 
\begin{equation}\label{eq_gaineps}
\mathit{G_{opt}}(l_c) \approx  \frac{1}{\epsilon}.
\end{equation}
In other words, the gain at the breakup location $L=l_c$ should be equal to $1/\epsilon$. Using equation \eqref{eq_gaineps}, we locate the gain in Fig. \ref{ResolventEps1}(b) for different forcing amplitudes $\epsilon=[10^{-2}-10^{-6}]$. At the given value of $\mathit{G_{h,fu}}$, a horizontal projection on the dominant frequency envelope will then decide the optimal forcing frequency for the given $\epsilon$. Finally, a vertical projection on $L$ from the intersection point on the dominant frequency envelope will provide the relevant breakup length $l_c$ for the forcing amplitude $\epsilon$. Extracting the results from Fig. \ref{ResolventEps1}(b), we compare the optimal forcing frequency and the breakup length for different $\epsilon$ with the nonlinear solutions of Section \ref{DNS_results} in Fig. \ref{ResolventEps2}. The close agreement between the two approaches shows the strength of the resolvent analysis in predicting the $\mathit{\omega_{opt}}$ and $l_c$ especially without any prior information from the nonlinear simulations. In Fig. \ref{ResolventEps2} the small difference in values in the two methods can likely be attributed to \textit{ad-hoc} definition of the required threshold for nonlinear effects to kick in and breakup to occur.  
\section{Response to white noise}\label{WhiteNoise}
\begin{figure}
\centering
\includegraphics[trim=100 0 100 0,clip,width=1\linewidth]{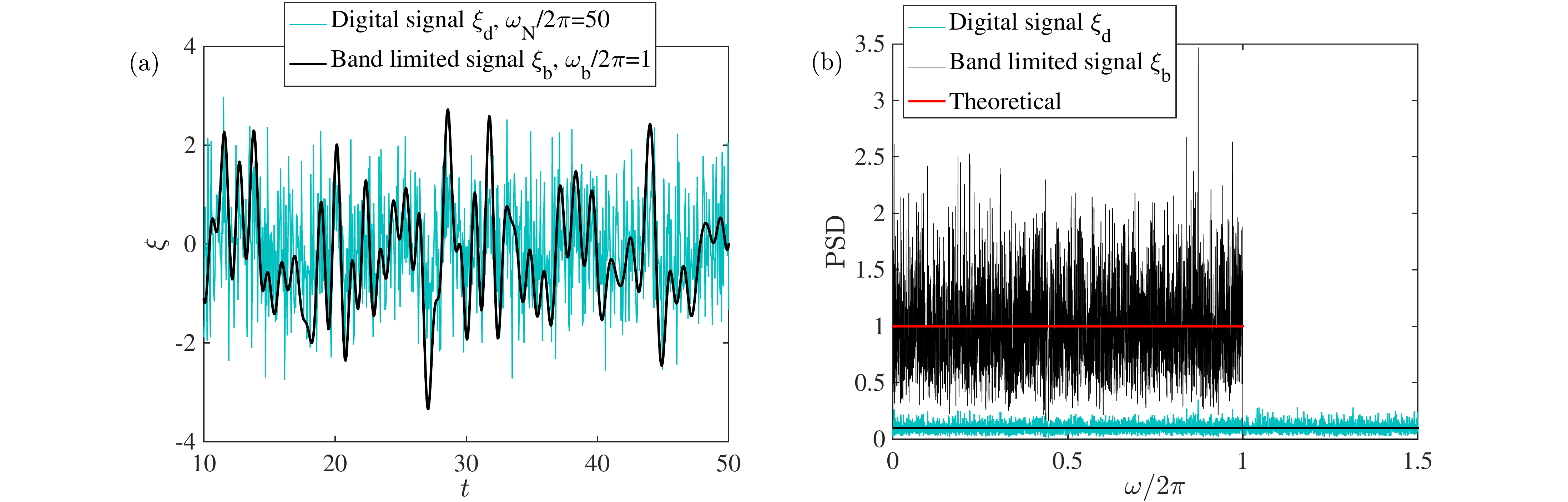}
\caption{(a) White noise signal with unit power, comparing a signal without filter and filtered using a band limiting frequency $\omega_b/2 \pi$. (b) Power spectral density comparison of these two signals with their theoretical value. The PSD is estimated using a Welch method in Matlab.}
\label{fig:WN1}
\end{figure}

Up to now, we were only interested in the response of the jet to an external disturbance characterised by a constant forcing frequency. However, in reality the external disturbance is more likely to be composed of a broadband frequency rather than being harmonic. Thus, to model this physical perturbation we carry out nonlinear simulations consistent with the scheme presented in Section \ref{chap3_dns} by exciting the jet at the nozzle by a white noise $\xi(t)$ defined in the time interval $[0~T]$ and formulated in a similar way as in \citet{mantivc2016saturation}. The white noise signal $\xi(t)$ is characterised by a constant power spectral density (PSD) $S_{\xi \xi}(\omega)=|\hat{\xi}(\omega)|^2$ where $\hat{\xi}(\omega)$ is the Fourier transform of $\xi(t)$  and has an infinite power $P$ defined as,
\begin{equation}\label{EqP}
P = \frac{1}{T}\int ^T_0 |\xi_T(t)|^2dt = \frac{1}{\pi}\int ^{\infty}_0 |\hat{\xi}|^2d\omega = \sigma ^2,
\end{equation}
where $\sigma$ is the variance. Even though a pure white noise has infinite power (as $S_{\xi\xi}(\omega) >0$), physical systems are usually characterised by a band-limited white noise. We thus filter the digital random signal $\xi_d(t)$ with a band limiting frequency $\omega_b/2 \pi=1$ to obtain the band limited white noise $\xi_b(t)$ as shown in Fig. \ref{fig:WN1}. For $\xi_d(t)$ the Nyquist frequency is set by $\omega_N/2 \pi$ which depends on the time step ($\delta t$) of the signal, such that $\omega_N/2 \pi=1/2\delta t$. Here we chose $\delta t=0.01$.
The noise $\xi_b(t)$ is normalised to have zero mean, unit variance and unit power, with a constant value for PSD, where $2|\hat{\xi_b}|^2=2 \pi/\omega_b$ which completely depends on the band limiting frequency. Finally we impose this filtered white noise as an inlet velocity condition for the jet defined by $\mathit{Oh_{in}}=0.3$, $\mathit{Bo_{in}}=0.1$ and $\mathit{We_{in}}=1.75$ and governed by the equations \eqref{eq3_govEq_dns} by replacing the boundary condition \eqref{eq3_dnsForcing} with,
\begin{equation}
\frac{du}{dt}\Bigg|_{(0,t)}=\epsilon\xi_b(t),
\end{equation}
where $\epsilon$ is the amplitude of the white noise signal. The forcing is applied at two different amplitudes $\epsilon=10^{-2}$ and $10^{-4}$ and for large times ($T=2000$) so as to achieve results which are time independent. For the MATLAB solver \texttt{ode23tb} with varying step size, the maximum time step size is set as $\delta t$ and white noise for intermediate time steps is obtained through interpolation. 

\begin{figure}
\centering
\includegraphics[trim=80 0 40 0,clip,width=1\linewidth]{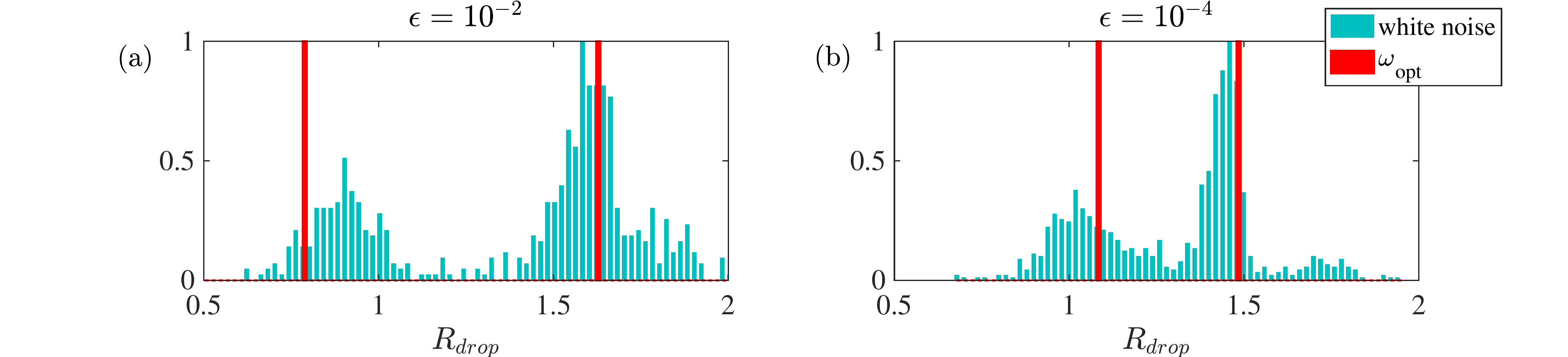}
\caption{Comparison of the normalised frequency of the drop radius $\mathit{R_{drop}}$ for jet defined by $\mathit{Oh_{in}}=0.3$, $\mathit{Bo_{in}}=0.1$ and $\mathit{We_{in}}=1.75$ and being forced at amplitude (a) $\epsilon=10^{-2}$ and (b) $\epsilon=10^{-4}$ by their respective optimal forcing frequency $\mathit{\omega_{opt}}$ (in red bars) and white noise (in cyan bars). For both the amplitudes, the white noise data is concentrated around two main drop sizes, representatives of the main and satellite drops. The most frequent drops sizes are $\mathit{R_{drop}} = 1.65$ and $1.45$ for $\epsilon=10^{-2}$ and $10^{-4}$ respectively. These values are close to the ones predicted by the nonlinear simulations of Fig. \ref{fig:WN1c} where the main drop size was predicted to be $1.62$ and $1.48$ for $\epsilon=10^{-2}$ and $10^{-4}$ respectively.}
\label{fig:WN4}
\end{figure}

\begin{figure}
\centering
\includegraphics[trim=80 0 40 0,clip,width=1\linewidth]{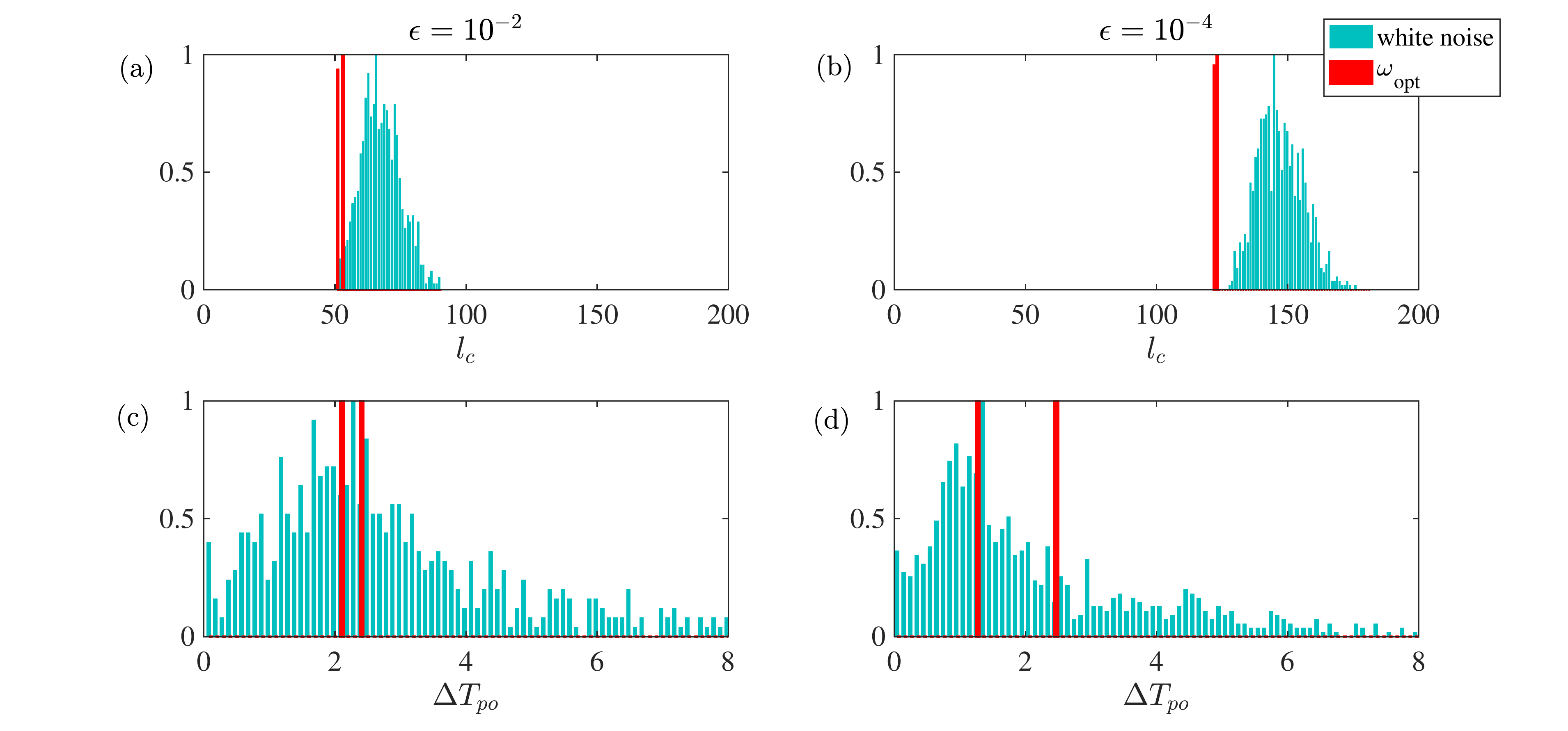}
\caption{ (a)-(b) refer to the comparison of the normalised frequency of the breakup length $l_c$ and (c)-(d) refer to the comparison of the normalised frequency of the breakup period $\mathit{\Delta T_{po}}$, each for a jet defined by $\mathit{Oh_{in}}=0.3$, $\mathit{Bo_{in}}=0.1$ and $\mathit{We_{in}}=1.75$. (a), (c) are subjected to a forcing amplitude $\epsilon=10^{-2}$ and (b), (d) to $\epsilon=10^{-4}$. The data in red corresponds to the optimal forcing frequency $\mathit{\omega_{opt}}$ and in cyan to the white noise. The most frequent white noise breakup length is close to the $l_c$ prediction in presence of the $\mathit{\omega_{opt}}$. For $\epsilon=10^{-2}$ and $10^{-4}$, the peak breakup period $\Delta T_{po}=2.45$ and $1.35$, respectively and is in close proximity to the results obtained from the nonlinear simulations of Fig. \ref{fig:WN1c}.}
\label{fig:WN5}
\end{figure}

At every pinch-off on the jet, we note the breakup length $l_c$, the pinch-off period $\mathit{\Delta T_{po}}$ and the drop radius $\mathit{R_{drop}}$ at the time of breakup. The distribution of the breakup characteristics is shown as as a histogram in Fig. \ref{fig:WN4} and \ref{fig:WN5} and compared with the expected response of the jet in presence of the pure $\mathit{\omega_{opt}(\epsilon)}$, which corresponds to $\mathit{\omega_{opt}}=1.38$ and $1.65$ for $\epsilon=10^{-2}$ and $10^{-4}$, respectively. The breakup characteristics for $\mathit{\omega_{opt}}$ have been discussed in Fig. \ref{fig:WN1c} and are depicted by red bars in Fig. \ref{fig:WN4} and \ref{fig:WN5}. 

The drop size distribution shown in Fig. \ref{fig:WN4}(a) highlights the two distribution peaks concentrated around $\approx 0.9$ and $\approx 1.65$, representing the group of satellite and main drops respectively. This behaviour also exists for smaller $\epsilon=10^{-4}$ where the radius is aggregated at $\approx 1.05$ and $\approx 1.45$. The results for the main drop size are coherent to the ones obtained in the presence of pure optimal forcing where $\mathit{R_{drop}}=1.62$ and $1.45$ for $\epsilon=10^{-2}$ and $10^{-4}$, respectively. Thus even in the presence of the white noise, the response of the jet is dominated by its expected behaviour at $\mathit{\omega_{opt}}$.

Unlike the drop radius, the peak of the distribution of breakup length obtained by imposing the white noise is not in close agreement with that of the optimal forcing as shown in Fig. \ref{fig:WN5}(a) and (b). Yet, we clearly see that the distribution spectrum shifts to large values of breakup length as $\epsilon$ is decreased, a behaviour similar to the one predicted by $\mathit{\omega_{opt}}$ where $l_c$ increases from $\approx 50$ to $\approx 125$ as $\epsilon$ is decreased. Similar conclusions can be drawn for the comparison of $\mathit{\Delta T_{po}}$ between white noise forcing and forcing with $\mathit{\omega_{opt}}$ from \ref{fig:WN5}(c) and (d) where we plot the breakup period between two consecutive drops.
\section{Conclusion and perspectives}\label{conc}
In this work, we inspect the response of a spatially varying gravitationally stretched jet subjected to an inlet velocity perturbation. The forcing is characterised through the frequency and the amplitude, the latter playing a major role in the determination of the optimal forcing frequency. The results of the numerical simulations performed on the nonlinear 1D Eggers \& Dupont equations shows an increase in optimal forcing frequency and the breakup length as the forcing amplitude is decreased. We found that the amplitude dependent preferred mode is a characteristic of gravity driven jets only. A pure capillary jet, base state of which is independent of gravity-induced stretching, does not sustain such a behaviour. In such cases, decreasing the forcing amplitude only resulted in an increase of the breakup length with the optimal frequency remaining fixed at all amplitudes. 

The linear stability theory characterised the jet flow used for nonlinear simulations as locally unstable and globally stable. Based on the absolute-convective transition criteria, we analysed the local stability at each section along the axial direction. The solution of the dispersion relation and the subsequent analysis for the downstream propagating spatial waves helped in confirming the predominant wave to be used for the zeroth order spatial gain expression. The strong non-parallelism of the base flow close to the nozzle motivated the incorporation of the WKBJ framework which markedly improved the prediction of the optimal forcing frequency in comparison to the resolvent analysis. However the spatial gain was still observed to be lower than the resolvent. As suggested by \citet{le2017capillary}, using advanced stability tools \citep{Schmid07} which accounts for non-parallel effects and non-modal growth leads to an estimation of a more realistic spatial response. 

This task was tackled using a resolvent analysis which accurately captured the linear response of stable jets in presence of an external forcing. Assuming a simple global amplitude breakup threshold criterion, the linear resolvent analysis becomes capable in predicting both the breakup length and the optimal forcing frequency given the amplitude of the forcing. The results of the nonlinear simulations and the resolvent for different forcing amplitudes are quantitatively comparable, thus underlining the importance of the resolvent analysis. Besides forcing the jet inlet with a fixed frequency, we also studied the response to a white noise, to analyse its natural response to a distributed forcing frequency range. Surprisingly, even in the presence of the white noise, the dominant response of the jet is close to the one seen from the optimal frequency at that amplitude. 

In presence of the external forcing, a dominant feature seen from the nonlinear simulations is the formation of a main and a satellite drop at the time of breakup. Nevertheless, to properly examine the consequence of the forcing amplitude on the final drop size, there is a need to enhance the nonlinear model by including the physics of drop coalescence and disintegration as done by \cite{driessen2011regularised}. Post breakup, the state of the jet after the pinch-off should be inferred from the system before the breakup. Additionally, the choice for drop curvature is of paramount importance since a given breakup can possess variety of drops shapes-each on different length scales \citep{kowalewski1996separation}.

On a different note, if the final aim is to eliminate the presence of satellite drops, the forcing should be modified such that it leads to the selective production of equisized drops. In this direction the work of \citet{chaudharyRed1980nonlinear}, who controlled satellite drops by forcing the jet with a suitable harmonic added to the fundamental; and \citet{driessen2014control}, who controlled the size of the droplet breaking off from a parallel jet by imposing a superposition of two Rayleigh-Plateau-unstable modes on the jet, could serve as the basis for formulating a theory for the spatially varying gravity jets. \\ \\ \\
I.S. thanks the Swiss National Science Foundation (grant no. 200021-159957). The authors would like to thank Eunok Yim for extremely valuable discussions on the modelling of white noise disturbance and in the interpretation of the local/global response. The authors would also like to thank Tobias Ansaldi and Giorgio Rocca who worked on the foundation of the numerical code for parallel jets as well as Adrien Jean Pierre Bressy for efficiently performing several simulations on the enhanced version of the numerical code provided to him. 
\section{Appendix}
\subsection{Numerical base state solution validation} \label{app_1g}
In this appendix, we show the validation of our numerically obtained base state solution of the governing equations \eqref{eq2_finalBS} with the experimental results of \cite{Rubio} for three different jet flows. The MATLAB \texttt{bvp4c} solver along with the boundary conditions stated in Section \ref{math_form} accurately captures the stretching (necking) close to the nozzle due to the effect of $\mathit{Bo_{in}}$.
\begin{figure}
	\centering
	\includegraphics[trim=50 0 20 0,clip,width=0.7\textwidth]{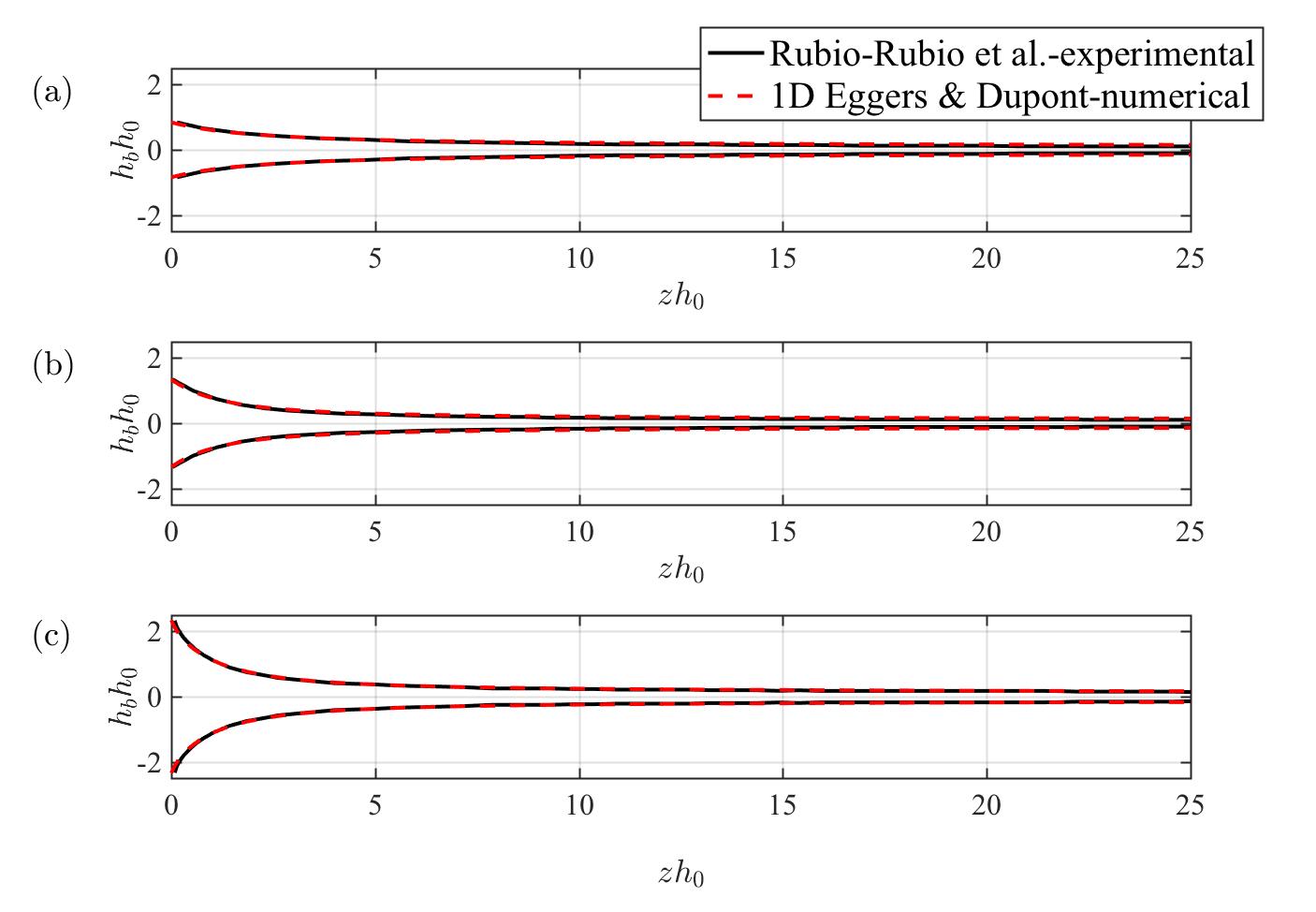}
	\caption{Comparison of the steady state solution with results from \cite{Rubio} for (a) $\mathit{Oh_{in}}=2.117$, $\mathit{We_{in}}=2.62\times10^{-2}$, $\mathit{Bo_{in}}=0.71$(b) $\mathit{Oh_{in}}=0.4799$, $\mathit{We_{in}}=6.06\times10^{-3}$, $\mathit{Bo_{in}}=1.81$ and (c) $\mathit{Oh_{in}}=0.7238$, $\mathit{We_{in}}=1.85\times10^{-3}$, $\mathit{Bo_{in}}=5.53$.}   
	\label{figBSvalid}
\end{figure}
\subsection{Effect of initial condition on breakup characteristics} \label{app_3g}
\begin{figure}
	\centering
	\includegraphics[trim=70 0 40 0,clip,width=1\textwidth]{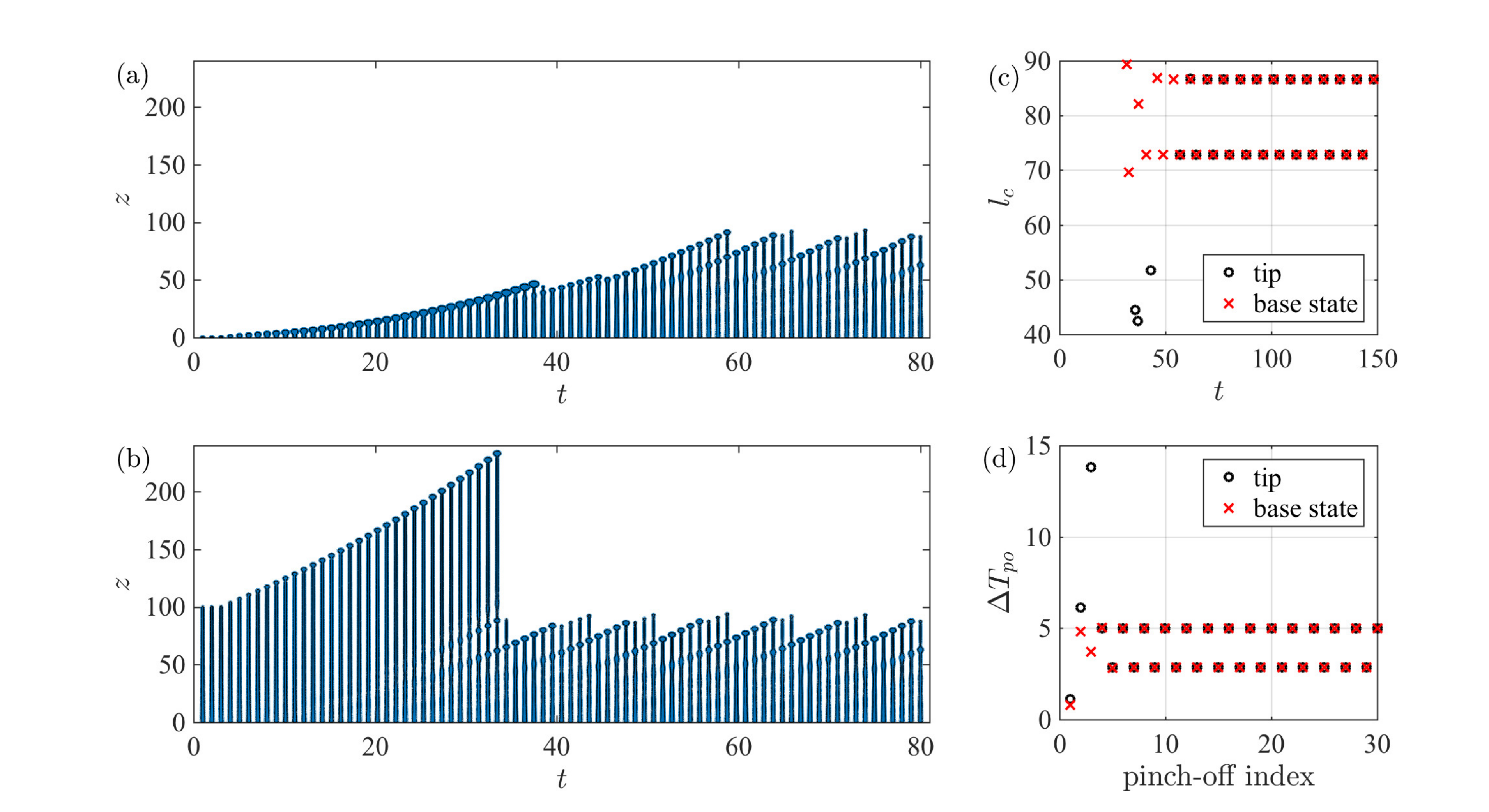}
	\caption{Time-sequence plot of a simulation with $\mathit{Oh_{in}}= 0.3$, $\mathit{We_{in}}= 1.75$, $\mathit{Bo_{in}} = 0.1$ excited with a forcing of amplitude $\epsilon=10^{-2}$ and frequency $\omega = 0.8$ and initialised (a) as a tip (b) using the base state solution. Comparison of the breakup length and period for both the cases is presented in (c) and (d), respectively.}   
	\label{tipVSbs}
\end{figure}

This section demonstrates the effect on breakup characteristics due to different initial conditions of the jet. Using the scheme described in Section \ref{Chap3_NumScheme} we perform numerical solutions for a jet with $\mathit{Oh_{in}}= 0.3$, $\mathit{We_{in}}= 1.75$, and $\mathit{Bo_{in}} = 0.1$ excited with a forcing of amplitude $\epsilon=10^{-2}$ and frequency $\omega = 0.8$.  In the first case, the jet is initialised as a circular tip of radius 1 (Fig. \ref{tipVSbs}(a)) and in the second case with the base state solution obtained by solving equation \eqref{eq2_finalBS} defined for an axial length of 100 (Fig. \ref{tipVSbs}(b)). In both the cases the numerical domain is considered large enough to capture all the breakups. As shown in Fig. \ref{tipVSbs}, both the jets with different initial conditions have different transient dynamics upto $t=55$ (tip) and $t=40$ (base state) after which they enter the permanent regime. In this regime, the breakup length $l_c$ and period $\mathit{\Delta T_{po}}$  are identical as shown in subplots (c) and  (d), respectively. It is thus safe to conclude that in the permanent regime the jet breakup is independent of the initial base state solution.
\subsection{Numerical scheme validation} \label{app_2g}
In this appendix, we show the validation of our numerical scheme described in Section \ref{Chap3_NumScheme} for the simulations of reduced 1D \cite{eggers1994} equations represented by the equation \eqref {eq3_govEq_dns}. For the purpose of validation, we use the numerical data of \citet{Hoeve2010} which are described for micro-jets of initial radius $h_0 = 18.5\, \mu$m with density $\rho=1098\,\mathrm{kg}/\mathrm{m}^3$,  viscosity $\eta = 3.65 \,$mPa.s, and surface tension $\gamma = 67.9 \,$mN/m. The jet is injected at a constant flow rate $Q = 0.35 \,$mL/min, corresponding to an initial jet velocity $U_0 = Q/(\pi h_0^2) = 5.4 \,$m/s. The flow can thus be described by the dimensionless numbers $\mathit{Oh_{in}}=0.1$ and $\mathit{We_{in}}=8.7$. Since the gravity effects are not considered in the experiment, we inject $\mathit{Bo_{in}}=0$ in equation \eqref {eq3_govEq_dns}. 
\begin{figure}
	\centering
	\includegraphics[trim=0 0 20 0,clip,width=0.75\textwidth]{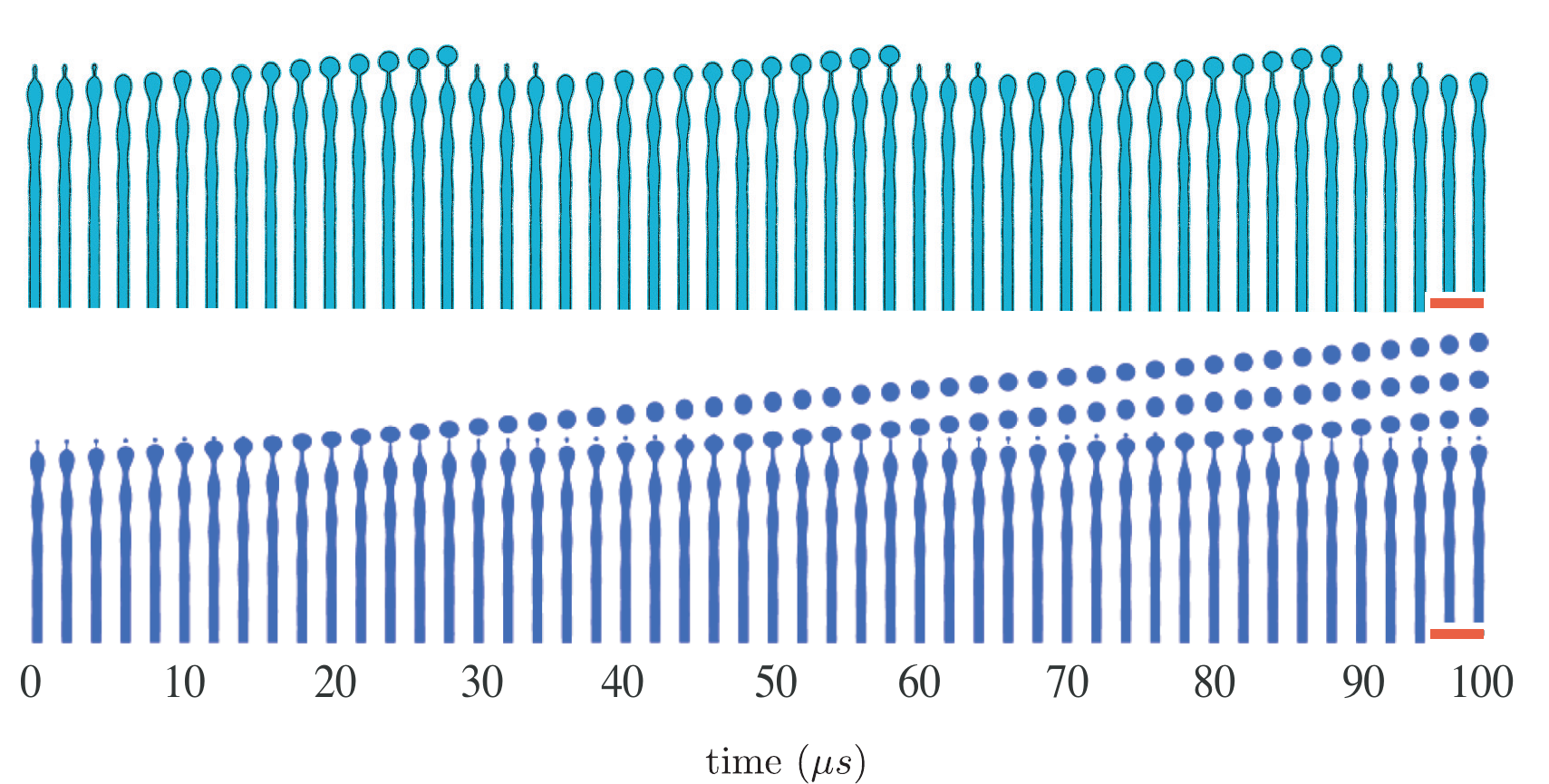}
	\caption{Numerical solutions of the governing equations \eqref{eq3_govEq_dns} for a jet in an inert medium with $\mathit{Oh_{in}}=0.1$, $\mathit{We_{in}}=8.7$ and $\mathit{Bo_{in}}=0$. (a) Results from our numerical scheme described in Section \ref{Chap3_NumScheme} and (b) experimentally-validated numerical results of \cite{Hoeve2010}. The red bar corresponds to a length scale of $200 \, \mu$m. }   
	\label{fighoeve}
\end{figure}
To initiate jet breakup in their numerical simulations, a harmonic modulation of the dimensional nozzle radius is applied as follows:
\begin{equation}
h(z=0,t) = h_0 + \delta \sin 2\pi nt,
\end{equation}
with $\delta/h_0 \approx 0.005$ the forcing amplitude, and $n$ the driving frequency. The latter is selected to match the optimum wavelength $\mathit{\lambda_{opt}}$ for jet breakup, that is, $n = U_0/\mathit{\lambda_{opt}}$. To ensure a constant flow rate $Q$ through the nozzle, the dimensional velocity is modulated correspondingly as
\begin{equation}
u_{0}(z=0,t) = \frac{h_0^2 U_0}{[h(z=0,t)]^2}.
\end{equation}
The amplitude of the wave imparted by the forcing at the nozzle grows until it equals the radius of the jet. Pinch-off or jet breakup is then defined as when the minimum width of the jet is below a predefined value set to $10^{-3} h_0$.

In our numerical simulations, we compute solutions to the governing equations \eqref{eq3_govEq_dns} with the same harmonic forcing and flow parameters as in \cite{Hoeve2010}. A hemispherical droplet described by $h=({h_0}^2 -z^2)^{1/2}$ is used as initial condition for the shape of the jet, the tip of which is therefore initially at $z=h_0$. The velocity is initialised to $u_0$ everywhere along the jet.  A fixed number of grid points, corresponding to a discretization size $dz = 0.05$, is uniformly distributed throughout the entire domain. The final validation is presented in Fig. \ref{fighoeve}, which shows comparison of the time series of the dynamics of jet breakup obtained from our numerical scheme and the numerical results from \cite{Hoeve2010}.

For both figures, the evolution of the jet shape is shown at time intervals of $2 \, \mu$s. Our numerical model predicts a breakup period of  $25 \, \mu$s and a breakup length of $ 856 \, \mu$m. The results of \cite{Hoeve2010}, have a breakup period of about $26$ to $30 \, \mu$s and a breakup length of about $800 \, \mu$m.  The error in breakup length between the two codes can be explained by the difference in grid size. Overall, Fig. \ref{fighoeve} shows a good agreement between both results and validates our numerical scheme.
\subsection{Comparison between resolvent and spatial analyses}\label{app_5g}
\begin{figure}
\centering
\includegraphics[trim=10 0 20 0,clip,width=0.6\linewidth]{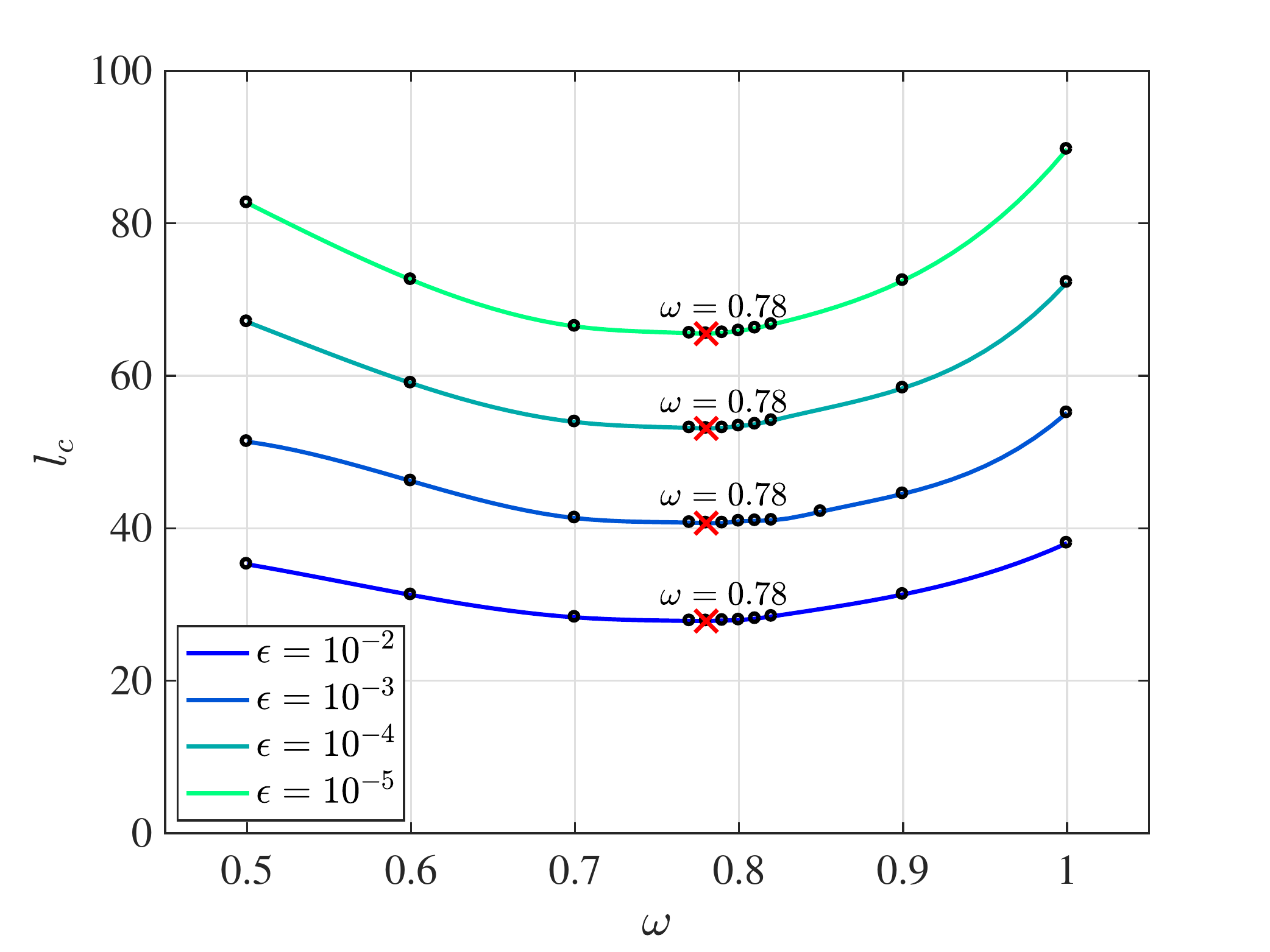}
\caption{Nonlinear simulation results for optimal forcing frequency $\mathit{\omega_{opt}}$ carried out for jet characteristics $\mathit{Oh_{in}}=0.3$, $\mathit{We_{in}}=1.75$, $\mathit{Bo_{in}}=0$. The breakup length $l_c$ is plotted as a function of forcing frequency $\omega$ for different forcing amplitudes $\epsilon$. } 
\label{dnsRes_noGravity}
\end{figure}

\begin{figure}
\centering
\includegraphics[trim=80 0 80 0,clip,width=1\linewidth]{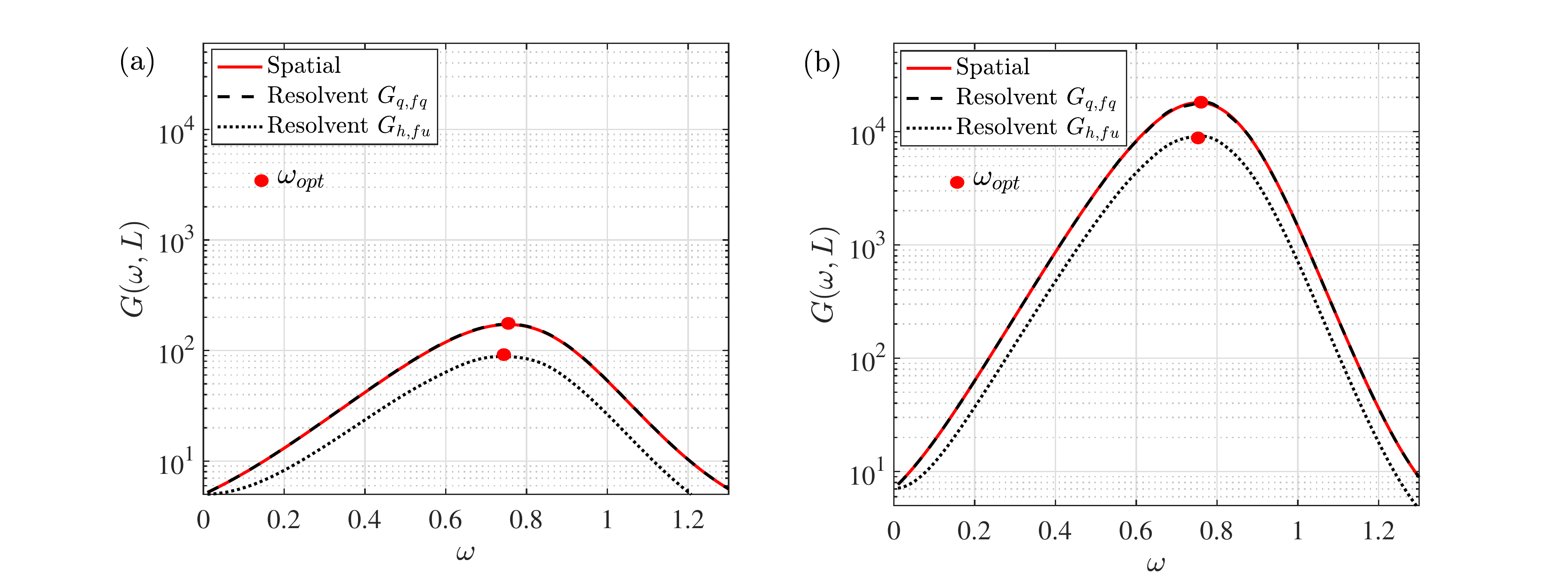}
\caption{Comparison of gain and $\mathit{\omega_{opt}}$ obtained from the resolvent analysis and spatial analysis for two different domain sizes (a) $L=25$ and (b) $L=50$ for a jet in absence of gravity and characterised by $\mathit{Oh}=0.3$ and $\mathit{We}=1.75$. Irrespective of the domain size and the method employed, the $\mathit{\omega_{opt}}$ lies between $[0.74~~0.75]$.} 
\label{Spat_gain}
\end{figure}

In this section we briefly show the preferred forcing frequency of the jet discussed in Section \ref{DNS_results} in the absence of gravity. The jet is characterised by $\mathit{Oh_{in}}=0.3$, $\mathit{We_{in}}=1.75$ and $\mathit{Bo_{in}}=0$. As shown in Fig. \ref{dnsRes_noGravity} nonlinear simulations for the governing equations \eqref{eq3_govEq_dns} for the zero gravity case using different forcing amplitudes $\epsilon$ show that $\mathit{\omega_{opt}}$ is independent of the chosen forcing amplitude, a behaviour in contrast to the situation where gravity is present (previously shown in Fig. \ref{dns-epsilon1}). 
Fig. \ref{Spat_gain} shows the comparison of gain, in absence of gravity, as function of forcing frequency using spatial and resolvent analysis for two different domain sizes. For convenience we also plot the resolvent gain $\mathit{G_{h,fu}}$ expressed in terms of the forcing applied for the nonlinear simulations. We note that all the curves, irrespective of the domain size, predict the same optimal forcing frequency $\mathit{\omega_{opt}}=0.74-0.76$, a value close to the nonlinear prediction of Fig. \ref{dnsRes_noGravity}. Moreover, unlike the situation with $Bo=0.1$, we notice that in the absence of gravity the magnitude of the gain at all frequencies is well captured by the spatial analysis. 
\subsection{Linear operator for eigenvalue problem}\label{app_6.1}
For the eigenvalue problem related to the global stability in Section \ref{global_stability}, the matrix $\mathit{M}$ is expressed as
\begin{equation}
\mathit{M}=  \begin{bmatrix}
  \mathit{M}_{11} &  \mathit{M}_{12} \\
  \mathit{M}_{21} &  \mathit{M}_{22}
                      \end{bmatrix}, 
\end{equation}
where the expressions  $\mathit{M}_{11} $, $\mathit{M}_{12} $, $\mathit{M}_{21} $ and $\mathit{M}_{22} $ denote the following differential equations:
\begin{subequations}\label{M14}
\begin{align}
\mathit{M}_{11}& =  -\frac{Q}{h^2_{b}}D + \frac{Qh'_{b}}{h^3_{b}}\rm I\!I,\\
\mathit{M}_{12}& =  -\frac{h_{b}}{2}D - h'_{b}\rm I\!I,\\
\mathit{M}_{21} &=   \sum_{k=1}^{4}s^{2k-1}\mathit{T}_k\ - 12{Oh_{in}}\: Q\bigg(\frac{h'^2_{b}}{h^2_{b}} + \frac{h''_{b}}{h_{b}}\bigg) D,\\
\mathit{M}_{22} &=3{Oh_{in}}\bigg( D^2 + \frac{2h'_{b}}{h_{b}D}\bigg)- \frac{Q}{h^2_{b}}D +\frac{2Qh'_{b}}{h^3_{b}}\rm I\!I.
\end{align}
\end{subequations}
In the group of equations \eqref{M14}, \rm I\!I is the identity operator, $D^n \equiv d^n/dz^n$, $s(z)=[1+(h'_{b})^2]^{-1/2}$ and 
\begin{subequations}
\begin{align}
\mathit{T}_{1} &=  \frac{1}{r^2_{b}}D - \frac{2r'_{b}}{r^3_{b}}\rm I\!I,\\
\mathit{T}_{2} &=  D^3 + \frac{h'_{b}}{h_{b}}D^2 - \bigg[\frac{(h'_{b})^2}{h^2_{b}} + \frac{h''_{b}}{h_{b}}\bigg] D - \frac{rh'_{b}h''_{b}}{h^2_{b}}\rm I\!I, \\
\mathit{T}_{3} &=   -6h'_{b}h''_{b}D^2 - 3 \bigg[\frac{(h'_{b})^2 h''_{b}}{h_{b}} + (h''_{b})^2 - {h'_{b}h'''_{b}}\bigg] D,\\
\mathit{T}_{4} &= 15(h''_{b})^2 (h'_{b})^2D.
\end{align}
\end{subequations}
\subsection{Global stability validation}\label{app_6.2}
\begin{figure}
	\centering
	\begin{subfigure}{1\textwidth} 
		\includegraphics[trim=40 0 0 0,clip,width=\textwidth]{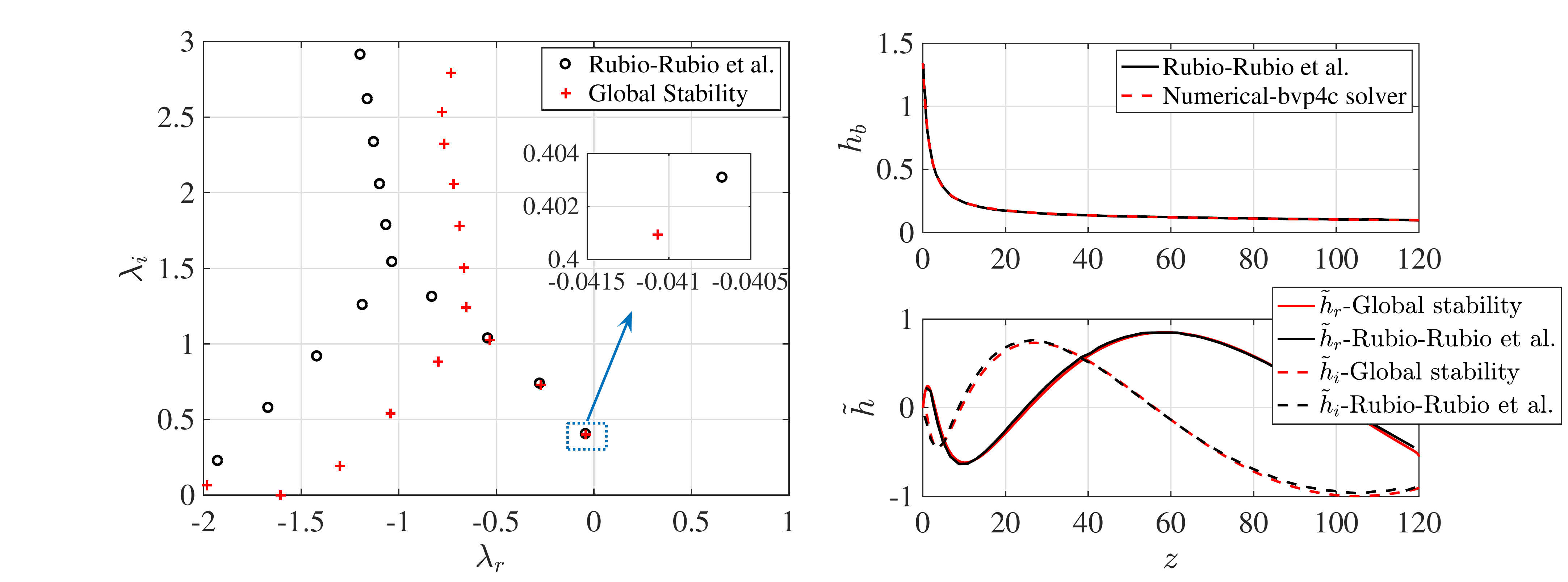}
		\caption{} \label{fig:Ng1a}
	\end{subfigure}
	\begin{subfigure}{1\textwidth} 
		\includegraphics[trim=40 0 0 0,clip,width=\textwidth]{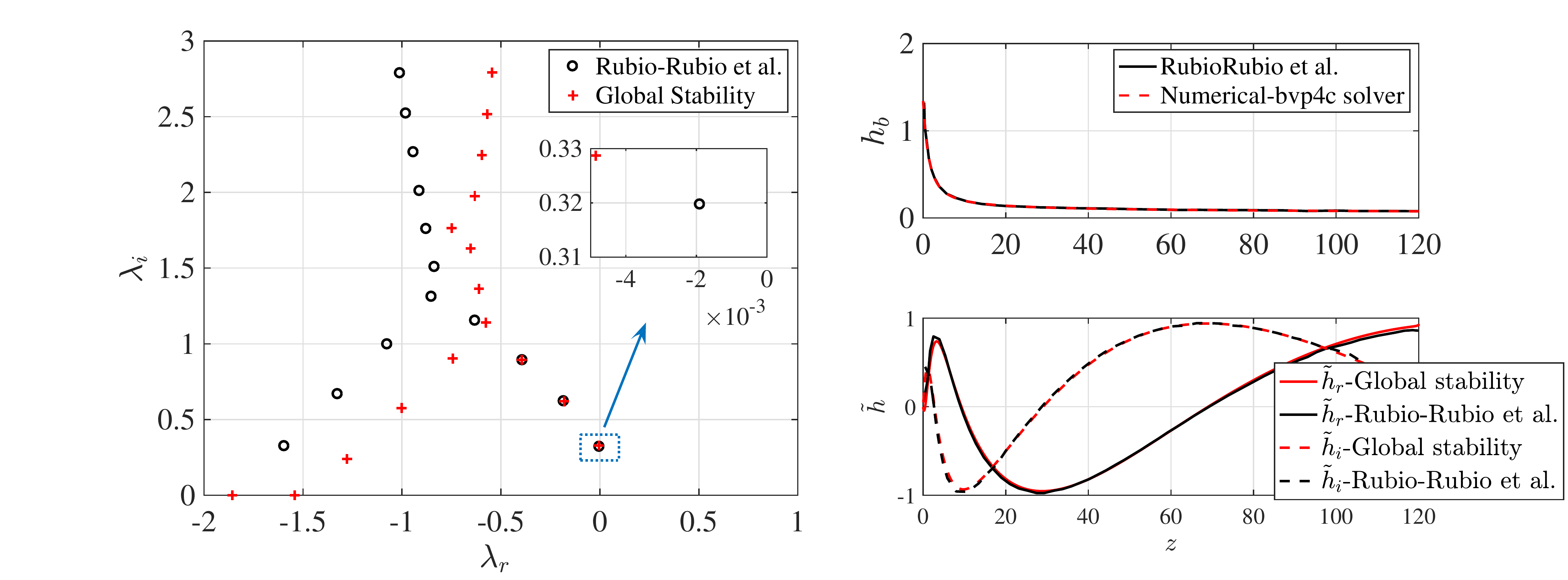}
		\caption{} \label{fig:Ng1b} 
			\end{subfigure}
		\caption{Eigenvalue spectrum $\lambda$, steady state shape of the jet $h_b$, and the real and imaginary parts of leading eigenfunction, $\tilde{h}$ for $\mathit{Oh_{in}}=1.68$, $\mathit{Bo_{in}}=1.81$ and (a) $\mathit{We_{in}}=8\times10^{-3}$ (b) $\mathit{We_{in}}=3\times10^{-3}$. Results in $black$ are from \cite{Rubio} and in $red$ are from the present stability model.}\label{fig:eig}
\end{figure}

\begin{figure}
	\centering
	\includegraphics[trim=0 0 0 0,clip,width=0.5\textwidth]{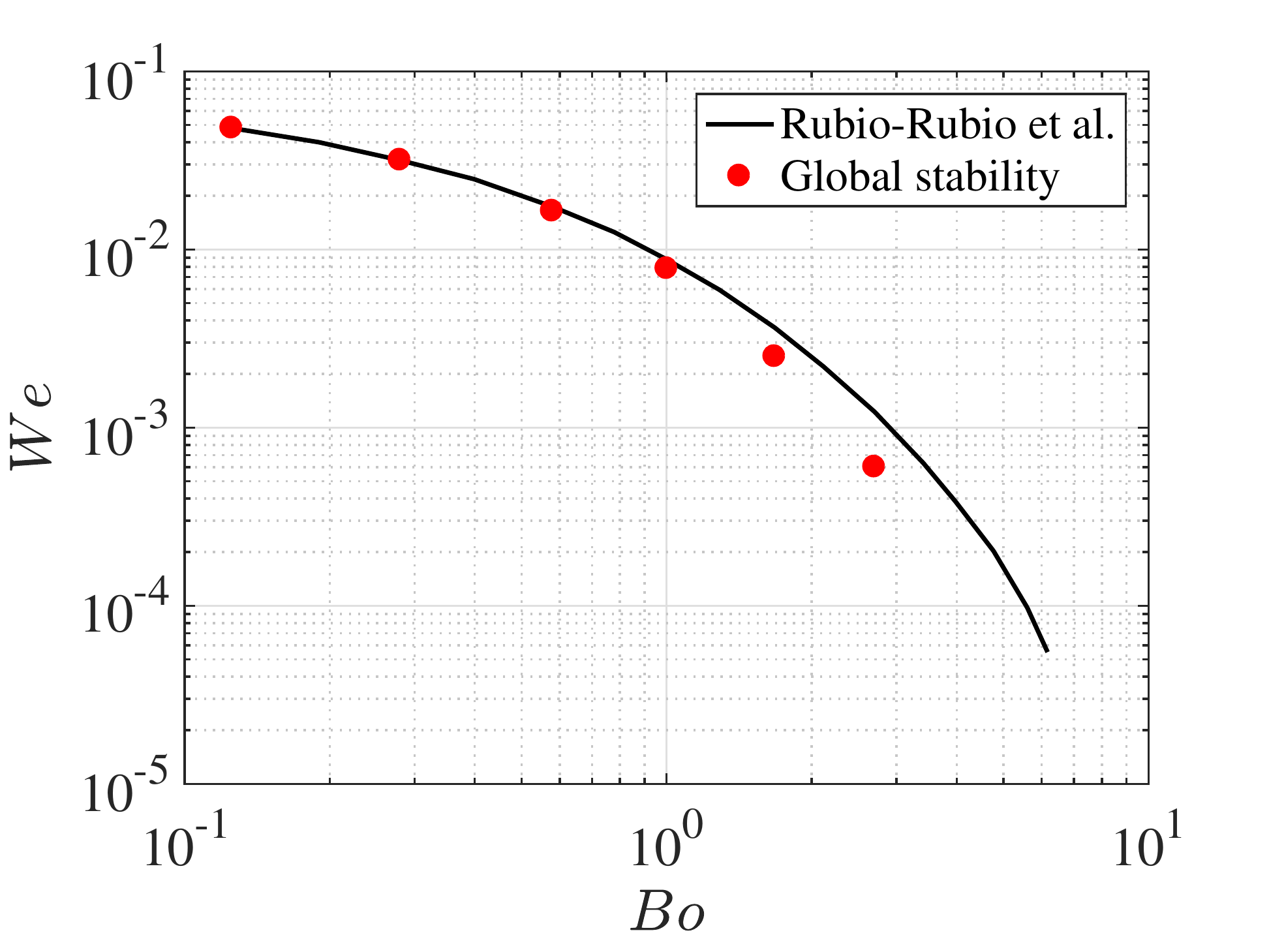}
	\caption{Comparison of critical Weber number, $\mathit{{We_{in}}_c}$.}
	\label{figsc}
\end{figure}

In this section, we present the validation of the numerical scheme used for the global stability analysis presented in Section \ref{global_stability}. The validation is done against the results of \cite{Rubio} where the stability analysis is based on the same 1D \cite{eggers1994} equations but made dimensionless using different characteristic length and time scales. The results of \cite{Rubio} are based on dimensionless numbers $\mathit{We_{in}}$, $\mathit{Bo_{in}}$ and Kapitza$(\Gamma)$. For the purpose of comparison we obtain the equivalent $\mathit{Oh_{in}}$ expressed as 
\begin{equation}
\mathit{Oh_{in}}= \frac{\Gamma}{3\mathit{Bo_{in}}^{0.25}}.
\end{equation}
For the eigenvalue problem, a non dimensional domain length $L=120$ is considered. For obtaining the dominant eigenvalue, the solution was computed for different values of  $N$ lying between $[150\,\, 250]$. \autoref{fig:eig} shows the validation of the eigenvalue spectrum with $N=200$, steady state and dominant eigenfunction for $\Gamma = 5.83$ and $\mathit{Bo_{in}} = 1.8$ and two different values of $\mathit{We_{in}}$. The results obtained from the present model are in good coherence with that obtained from \cite{Rubio}. Comparing Fig. \ref{fig:eig}(a) and (b), we can observe there is critical Weber number $\mathit{{We_{in}}_c}$, for which the jet becomes marginally unstable as the real part of the leading eigenvalue is slightly positive. For the given values of $\mathit{Bo_{in}}$ and $\mathit{Oh_{in}}$, the $\mathit{{We_{in}}_c}$ is therefore equal to $3\times10^{-3}$. If the Ohnesorge number is fixed and the bond number is varied, we can get the corresponding $\mathit{{We_{in}}_c}$ for each value of $\mathit{Bo_{in}}$.  \autoref{figsc} represents the curve for the $\mathit{{We_{in}}_c}$, below which the jet becomes linearly unstable. The results are compared to those obtained from \cite{Rubio}. 
\subsection{WKBJ formulation for axisymmetric 1D Eggers \& Dupont equations}\label{wkbj_all}
\subsubsection{Linearised equations}
Considering linear perturbations ($a^\prime,u^\prime$) in jet interface and velocity around the base flow ($a_b,u_b$), the linearised system of equations is written as:
\begin{subequations}\label{Linear_h2}
\begin{align}
\frac{\partial a^\prime}{\partial t} =& (-A_1-A_2D_1) a^\prime + (-A_3-A_4D_1)u^\prime , \\
\frac{\partial u^\prime}{\partial t} =& (-B_1-B_2D_1-B_3D_2-B_4D_3)a^\prime + (-B_1-B_2D_1-B_3D_2)u^\prime .
\end{align}
\end{subequations} 
where  $D_i$ with $i=1..3$ are the differential operators with respect to $z$.
Equation \eqref{Linear_h2} can be reformed as
\begin{equation}\label{}
[\mathbf{\dot{q^\prime}}] = \mathit{K}  [\mathbf{q^\prime}],
 \end{equation}
 where 
 \begin{equation}\label{r}
 [\mathbf{{q^\prime}}]  =  \begin{bmatrix} {a^\prime} \\ {u^\prime}\end{bmatrix}, ~~~~~~ \mathit{K} 
 = 
 \begin{bmatrix}
   -A_1-A_2D_1       &  -A_3-A_4D_1    \\
   -B_1-B_2D_1-B_3D_2-B_4D_3       &   -B_5-B_6D_1-B_7D_2 \\
\end{bmatrix},\\
 \end{equation}
where the coefficients $A_{1..4}$ and $B_{1..7}$ are given as,
\begin{subequations}
\begin{align}
A_1 =& {u_b}'(z),  \\
A_2= &  {u_b}(z), \\ 
A_3 =& {a_b}'(z),  \\
A_4 =& {a_b}(z),  \\
%
B_1 = & \frac{15 \left(-2 {a_b}'(z)^3-4 {a_b}(z) {a_b}'(z)+{a_b}'(z)^3 \left(-{a_b}''(z)\right)+{a_b}(z) {a_b}'(z) {a_b}''(z)^2\right)}{16 S^{7/2}}  \nonumber \\ 
& -\frac{3 \left(2 {a_b}(z) {a_b}^{(3)}(z)-8 {a_b}'(z)+{a_b}'(z) {a_b}''(z)^2-2 {a_b}'(z) {a_b}''(z)\right)}{8 S^{5/2}} \nonumber \\
& +  \frac{{a_b}^{(3)}(z)}{2 S^{3/2}} - \frac{3 \left({Oh_{in}} {a_b}'(z) {u_b}'(z)\right)}{{a_b}(z)^2},  \\
B_2 = & \frac{15 \left(-2 {a_b}'(z)^4-4 {a_b}(z) {a_b}'(z)^2+{a_b}'(z)^4 \left(-{a_b}''(z)\right)+{a_b}(z) {a_b}'(z)^2 {a_b}''(z)^2\right)}{32 S^{7/2}}  \nonumber \\
& -\frac{3 \left({a_b}(z) {a_b}''(z)^2-8 {a_b}'(z)^2+{a_b}(z) {a_b}^{(3)}(z) {a_b}'(z)-4 {a_b}'(z)^2 {a_b}''(z)-4{a_b}(z)\right)}{8 S^{5/2}} \nonumber \\
& - \frac{{(a_b}''(z)+2)}{2S^{3/2}} \nonumber \\
& + \frac{3Oh{u_b}'(z)}{a_b(z)},\\
B_3 = & \frac{3 \left({a_b}'(z)^3-2 {a_b}(z) {a_b}'(z) {a_b}''(z)\right)}{8 S^{5/2}}--\frac{{a_b}'(z)}{2 S^{3/2}},  \\
B_4 = & \frac{{a_b}(z)}{2 S^{3/2}}, \\
B_5 =& -{u_b}'(z),  \\
B_6 =& \frac{3 {Oh_{in}} {a_b}'(z)}{{a_b}(z)}-{u_b}(z),  \\
B_7 =& 3Oh.   
\end{align}
\end{subequations}
In the above equation $S$ replaces the term $\frac{1}{4} {a_b}'(z)^2+{a_b}(z)$. 
\subsubsection{Linearised equation expressed in terms of slow variable }
For the WKB ananlysis, we then introduce the spatial scales. The fast spatial scale $z$ is replaced by the slow scale $Z$, such that $Z=\eta z$. The base flow is now expressed as a function of $Z$ such that $a_b(Z)$ and $u_b(Z)$. Let us consider the following normal mode expansion for the perturbation:
\begin{equation}\label{}
\mathbf{q^\prime}(Z,t) = \mathbf{\hat q}(Z)  \exp\Big[ \text{i} \Big( \frac{1}{\eta} \int_{0}^{Z} k(Z',\omega)dZ' -\omega t \Big) \Big]. 
 \end{equation}
Injecting the transformations \eqref{wkbj_transf} into \eqref{Linear_h2}, the linearised equations on a weakly non-parallel baseflow equations are expressed through equations \eqref{linear_wkbj_mass}-\eqref{linear_wkbj_mom},
\begin{subequations}
\begin{align*}
\frac{\partial}{\partial t} &\rightarrow-\text{i}\omega,  \\
\frac{\partial}{\partial z} &\rightarrow \phantom{-}\text{i}k+\eta \frac{\partial}{\partial Z}, \\ 
\frac{\partial^2}{\partial z^2} &\rightarrow -k^2 + \text{i}\eta \Bigg(\frac{\partial k}{\partial Z} + 2k\frac{\partial }{\partial Z}\Bigg)+ \eta ^2 \frac{\partial ^2}{\partial Z^2},  \\
\frac{\partial^3}{\partial z^3} &\rightarrow -\text{i}k^3 - 3\eta k \Bigg(\frac{\partial k}{\partial Z}+ k\frac{\partial}{\partial Z} \Bigg) + \eta^2 \Bigg( 3\text{i} \frac{\partial k}{\partial Z}\frac{\partial}{\partial Z} +  3\text{i}k \frac{\partial^2}{\partial Z^2} + \text{i}\frac{\partial^2 k}{\partial Z^2} \Bigg)  +\eta^3 \frac{\partial ^3}{\partial Z^3}.
\refstepcounter{equation}\tag{\theequation -d}\label{wkbj_transf}
\end{align*}
\end{subequations}
where the continuity equation converts to:
\begin{equation}\label{linear_wkbj_mass}
(-\text{i}\omega + \text{i}ku_b)\hat{a} + (\text{i}ka_b)\hat{u}=-\eta\Bigg[ \frac{\partial}{\partial Z}(a_b\hat{u}+u_b\hat{a})\Bigg],
\end{equation}
and the momentum equation transforms as,
\begin{equation}\label{linear_wkbj_mom}
\begin{aligned}
\text{i}\omega \hat{u}=&-\eta\frac{\partial u_b}{\partial Z}\hat{u} - \Bigg( u_b-\eta \frac{3Oh}{a_b}\frac{\partial a_b}{\partial Z} \Bigg)\Bigg(\eta \frac{\partial}{\partial Z} + \text{i} k \Bigg)\hat{u} \\
& +3Oh\Bigg(-k^2 + \eta \text{i}\Bigg(\frac{\partial k}{\partial Z} +2k\frac{\partial}{\partial Z} \Bigg) + \eta^2 \frac{\partial ^2}{\partial Z^2} \Bigg)\hat{u}-\Bigg(\eta\frac{3}{4a_b^{5/2}}\frac{\partial a_b}{\partial Z}\Bigg)\hat{a}\\
&+\Bigg(\frac{1}{2a_b^{3/2}} + \frac{3\eta Oh}{a_b}\frac{\partial u_b}{\partial Z}\Bigg)\Bigg(\eta \frac{\partial}{\partial Z}+\text{i} k \Bigg)\hat{a}\\
&-\frac{\eta}{2a_b^{3/2}}\frac{\partial a_b}{\partial Z}\Bigg(-k^2 + \eta \text{i} \Bigg( \frac{\partial k}{\partial Z} + 2k\frac{\partial}{\partial Z}\Bigg) +\eta^2 \frac {\partial ^2}{\partial Z^2} \Bigg)\hat{a}\\
&+ \frac{1}{2a_b^{1/2}}\Bigg(-\text{i}k^3 -3\eta k \Bigg( \frac{\partial k}{\partial Z} +k\frac{\partial}{\partial Z} \Bigg) + \mathcal{O}(\eta^2)\Bigg)\hat{a}
\end{aligned}
\end{equation}
Defining  $\mathbf{\hat q}^{(1)}=[\hat a^{(1)} ~\hat u^{(1)}]$ and $\mathbf{\hat q}^{(2)}=[\hat a^{(2)} ~\hat u^{(2)}]$, we now consider the asymptotic expansion:
\begin{equation}\label{}
\mathbf{\hat q} \sim A(Z)\mathbf{\hat q}^{(1)}(Z) + \eta \mathbf{\hat q}^{(2)}(Z) + \cdots,
 \end{equation}
and inject it into the governing equations \eqref{linear_wkbj_mass}-\eqref{linear_wkbj_mom} to obtain the local stability problem at $\eta^0$ and $\eta^1$. \\ \\
\textbf{\underline{Order $\eta^0$}}
At zeroth-order in $\eta$, the local stability problem is retrieved:
\begin{subequations}\label{linear_e0}
\begin{align}
\underbrace{(-\text{i}\omega+\text{i}ku_b)}_{\mathit{L} _{11}}&\hat a^{(1)} + \underbrace{(\text{i}ka_b)}_{\mathit{L} _{12}}\hat u^{(1)} =0, \\
\underbrace{-\frac{\text{i}k}{2a_b^{3/2}}(1-k^2a_b)}_{\mathit{L} _{21}}&\hat a^{(1)}+\underbrace{\addstackgap[11pt](-\text{i}\omega+\text{i}ku_b +3Ohk^2)}_{\mathit{L} _{22}}\hat u^{(1)}=0.
\end{align}
\end{subequations}
The system of equations represented by \eqref{linear_e0} can be reframed using the linear operator $\mathit{L}$, such that
\begin{equation}\label{line-eigen}
 \mathit{L}[\mathbf{\hat q}^{(1)}] =0, \phantom{------}\text{where the linear operator} \phantom{-} \mathit{L}  = 
 \begin{bmatrix}
     \mathit{L} _{11}       &   \mathit{L} _{12}    \\
     \mathit{L} _{21}       &   \mathit{L} _{22}   \\
\end{bmatrix}.\\
\end{equation}
Substituting expression for $\hat u^{(1)}$ from \eqref{linear_e0}a into \eqref{linear_e0}b, we finally obtain:
\begin{equation}\label{linear_e0_kpoly}
\Bigg(-\omega^2 +2u_b\omega k +\Big(-\frac{1}{2 \sqrt{a_b}}-u_b^2 -3\text{i}Oh\omega\Big)k^2 +3\text{i}Ohu_b k^3 +\frac{\sqrt{a_b}}{2}k^4\Bigg)\hat a^{(1)} =0,
\end{equation}
the solution of which gives the four roots of $k$ for a given $\omega$. The relevant $k$ branch is tracked as discussed in Section \ref{LinStab}. For a given $\omega$ and a predetermined $k$, the solution of the linear problem \eqref{line-eigen} gives the response $\mathbf{\hat q}^{(1)}$, a parameter needed to solve the local stability problem at $\eta^1$. \\ \\
\textbf{\underline{Order $\eta^1$}}: At first order we obtain,
\begin{equation}\label{}
 \mathit{L}[\mathbf{\hat q}^{(2)}] =  \mathit{Q}[A\mathbf{\hat q}^{(1)}], 
\end{equation}
where operator $\mathit{Q}$ can be split into two parts:
\begin{equation}\label{}
\mathit{Q}[A\mathbf{\hat q}^{(1)}] = \mathit{R}[\mathbf{\hat q}^{(1)}]\frac{\text{d}A}{\text{dZ}} + \mathit{S}[\mathbf{\hat q}^{(1)}] A.
\end{equation}
Operator $ \mathit{R}$ is expressed as:
\begin{gather}
 R= \begin{bmatrix}
    -u_b &
   -a_b \\
   \Bigg(\frac{1-3k^2a_b}{2a_b^{3/2}}\Bigg)  &
  (6\text{i}Ohk-u_b) 
   \end{bmatrix},
\end{gather}
and $\mathit{S}$ is defined as,
\begin{equation*}
\mathit{S} = 
  \begin{pmatrix*}[l]
       s_{11} & s_{12}   \\
     s_{21} & s_{22}
  \end{pmatrix*},
\end{equation*}
where the individual parameters are expressed as:
\begin{subequations} 
\begin{alignat*}{3}
&s_{11} =-\frac{\partial u_b}{\partial Z}-u_b\frac{\partial}{\partial Z},\\
&s_{12} = -\frac{\partial a_b}{\partial Z}-a_b\frac{\partial}{\partial Z}, \\
&s_{21} =  \phantom{-}\frac{3\text{i}Ohk}{a_b}  \frac{\partial u_b}{\partial Z}-\frac{1}{4a_b^{5/2}}\Bigg((3-2k^2a_b)\frac{\partial a_b}{\partial Z} +6ka_b^2 \frac{\partial k}{\partial Z} +2a_b(-1+3k^2a_b)\frac{\partial}{\partial Z}\Bigg),\\    
&s_{22} =  \phantom{-}3\text{i}Oh\Bigg(\frac{k}{a_b}\frac{\partial a_b}{\partial Z}+\frac{\partial k}{\partial Z} +2k\frac{\partial}{\partial Z}\Bigg) -u_b \frac{\partial }{\partial Z}-\frac{\partial u_b}{\partial Z}.\\
             \end{alignat*}
\end{subequations}
As explained in \citealp{huerre1998hydrodynamic,viola2016mode}, in order to have solutions of the inhomogeneous equation $\mathit{L}[\mathbf{\hat q}^{(2)}] =  \mathit{Q}[A\mathbf{\hat q}^{(1)}] $, the forcing term $Q$ should be in the image of the operator $L$. This implies that $Q$ should be orthogonal to the corresponding adjoint eigenfunction $\mathbf{\tilde q}^{(1)}$ of the adjoint operator $\tilde{L}$, with respect to the defined inner product,
\begin{equation}
\underbrace{\addstackgap[11pt]R[\mathbf{\hat{q}}^{(1)}]\mathbf{\tilde{q}}^{(1)} \frac{\mathrm{d}A}{\mathrm{d}Z}}_{M(Z)} + \underbrace{\addstackgap[11pt]S[\mathbf{\hat{q}}^{(1)}]\mathbf{\tilde{q}}^{(1)}A}_{N(Z)} =
L[\mathbf{\hat{q}}^{(2)}]\mathbf{\tilde{q}}^{(1)} = \mathbf{\hat{q}}^{(2)} \tilde{L}[\mathbf{\tilde{q}}^{(1)}] =0.
\end{equation}
This leads to the amplitude equation,
\begin{equation}
M(Z)\frac{\mathrm{d}A}{\mathrm{d}Z} +N(Z) A = 0,
\end{equation}
solving which we obtain the amplitude solution $A(Z)$ which should then be expressed in terms of the fast length scale $z$. Finally, at first order, the response if given by,
\begin{equation}
\mathbf{q}^\prime(z)\sim A(\eta z) \mathbf{\hat{q}}^{(1)}(z) \exp \Bigg(\int_0^z -k_i(z)^\prime \mathrm{d}z^\prime) \Bigg)\exp\Bigg[ \text{i} \Bigg(\int_0^zk_r(z^\prime)\mathrm{d}z^\prime -\omega t\Bigg)\Bigg] .
\end{equation}
\bibliographystyle{jfm}
\bibliography{gravity}

\end{document}